\documentclass[10pt]{article}

\usepackage{graphicx}
\usepackage{float}
\usepackage[abs]{overpic}
\usepackage{rotating}

\usepackage[scriptsize, figurename=FIG., tablename=TAB.]{caption}
\usepackage{subfig}
\usepackage{pstricks}
\usepackage{pstricks-add}
\usepackage{pst-plot}
\usepackage{color}
\usepackage{enumitem}
\usepackage{chngcntr}
\usepackage[all]{xy}
\usepackage[multidot]{grffile}
\usepackage{sectsty}
\usepackage{titlecaps}
\usepackage[colorlinks=true]{hyperref}

\usepackage{amsmath,amsfonts,amssymb,amsmath,mathrsfs,latexsym,bbm,dsfont,amsthm,empheq}
\usepackage{relsize,exscale}

\let\oldsqrt\sqrt
\def\sqrt{\mathpalette\DHLhksqrt}
\def\DHLhksqrt#1#2{%
\setbox0=\hbox{$#1\oldsqrt{#2\,}$}\dimen0=\ht0
\advance\dimen0-0.2\ht0
\setbox2=\hbox{\vrule height\ht0 depth -\dimen0}%
{\box0\lower0.4pt\box2}}

\newcommand{\RN}[1]{%
  \textbf{\textup{\uppercase\expandafter{\romannumeral#1}}}%
}

\DeclareMathOperator{\Lagr}{\mathcal{L}}

\sectionfont{\MakeUppercase}
\subsubsectionfont{\bfseries\itshape}

\def\equationautorefname~#1\null{Eq.~(#1)\null}

\hypersetup{allcolors=[rgb]{0.19,0.17,0.57}}

\begin{document}

\pagenumbering{arabic}	
\title{Geodesic motion around traversable wormholes supported by a massless conformally-coupled scalar field}
\author{Felix Willenborg, Saskia Grunau, Burkhard Kleihaus, Jutta Kunz \\
\vspace{0.1cm} \\
Institut f\"ur Physik, Universit\"at Oldenburg, D--26111 Oldenburg, Germany \\
}
\date\today
\maketitle

\begin{abstract}
We consider a traversable wormhole solution of Einstein's gravity conformally coupled to a massless scalar field, a solution derived by Barcelo and Visser based on the Janis-Newman-Winicour-Wyman spacetime. We study the geodesic motion of timelike, lightlike and spacelike particles in this spacetime. We solve the equations of motion analytically in terms of the Weierstra\ss \ functions and discuss all possible orbit types and their parameter dependence. Interestingly, bound orbits occur for timelike geodesics only in one of the two worlds. Moreover, under no conditions there exist timelike two world bound orbits.
\end{abstract}

\section{Introduction}
\label{Intro:Sec}
Since the concept of wormholes was introduced by Einstein and Rosen in 1935 \cite{Einstein:1935tc}, wormholes have been discussed in a widely spread manner in the literature. Of particular interest are Lorentzian traversable wormholes (see e.g.~\cite{Morris:1988cz,Visser:1995cc,Lobo:2017eum}). In recent years their potential astrophysical signatures have received much attention. For instance, gravitational lensing by wormholes has been studied \cite{Cramer:1994qj,Safonova:2001vz,Perlick:2003vg,Nandi:2006ds,Nakajima:2012pu,Tsukamoto:2012xs,Kuhfittig:2013hva,Tsukamoto:2016zdu}, wormhole shadows have been investigated \cite{Bambi:2013nla,Nedkova:2013msa,Ohgami:2015nra,Shaikh:2018kfv}, the accretion disks surrounding wormholes have been considered\cite{Harko:2008vy,Harko:2009xf,Bambi:2013jda,Zhou:2016koy,Lamy:2018zvj}, and the viability of wormholes as black hole alternatives has been addressed (see e.g. \cite{Damour:2007ap,Bambi:2013nla,Azreg-Ainou:2014dwa,Dzhunushaliev:2016ylj,Cardoso:2016rao,Konoplya:2016hmd,Nandi:2016uzg,Bueno:2017hyj}). Moreover, first astrophysical searches for wormholes have been carried out \cite{Abe:2010ap,Toki:2011zu,Takahashi:2013jqa}.

The construction of traversable wormholes requires the violation of the \textit{Null Energy Condition} (NEC), which can, for instance, be achieved by the presence of exotic matter, yielding the Ellis (or Bronnikov-Ellis) wormholes of General Relativity \cite{Ellis:1973yv,Ellis:1979bh,Bronnikov:1973fh,Kashargin:2007mm,Kashargin:2008pk,Kleihaus:2014dla,Chew:2016epf}. On the other hand, no such exotic matter is needed to obtain traversable wormholes in many modified theories of gravity (see e.g. \cite{Hochberg:1990is,Fukutaka:1989zb,Ghoroku:1992tz,Furey:2004rq,Lobo:2009ip,Bronnikov:2009az,Kanti:2011jz,Kanti:2011yv,Harko:2013yb}).

Staying with General Relativity,  Barcelo and Visser \cite{Barcelo:1999hq}, however, showed, that the NEC can also be violated and traversable wormhole solutions be found, when General Relativity is conformally coupled to a massless scalar field, thus avoiding the presence of exotic matter while retaining General Relativity for the description of gravity. Indeed, by employing the {\sl  new improved energy-momentum tensor $T_{\mu\nu}$} of Callan \textit{et al.}~\cite{Callan:1970ze}, they obtained an interesting set of wormhole solutions. As pointed out there \cite{Callan:1970ze,Barcelo:1999hq}, this new improved energy-momentum tensor has the same set of Poincaré generators as the one of a minimally coupled scalar field, while its matrix elements are finite in every order of renormalized perturbation theory, rendering the resulting theory rather attractive.

To obtain the set of static spherically symmetric wormhole solutions of this theory, Barcelo and Visser started from the well known solution found by Janis, Newman and Winicour and independently by Wyman \cite{Janis:1968zz,Wyman:1981bd,Virbhadra:1997ie}, the JNWW solution, looking for solutions of the new set of field equations, which are conformally related to the known JNWW solution. This way they obtained a differential equation for the conformal factor, to be solved together with the scalar field equation. The resulting set of solutions then represented a generalization of those of Froyland \cite{Froyland:1981yd}, and Agnese and La Camera \cite{Agnese:1985xj}.

Here we focus on the traversable wormhole solutions which form an intriguing subset of the full set of solutions found by Barcelo and Visser \cite{Barcelo:1999hq}. Let us emphasize, that these new wormholes are based on General Relativity and do not need any exotic matter. Their massless scalar field is an ordinary scalar field with a conventional kinetic term in the Lagrangian. Instead, the existence of these wormholes is based on the conformal coupling of the scalar field to gravity, which allows for a violation of the NEC. Being based on an attractive underlying theory makes these wormholes interesting objects to study their physical properties.

The absence of exotic matter may have another big advantage, namely the absence of the notorious radial instability of wormholes in General Relativity supported by a phantom scalar field \cite{Shinkai:2002gv,Gonzalez:2008wd,Gonzalez:2008xk,Torii:2013xba,Dzhunushaliev:2013jja}. Clearly, a study of the stability of these wormhole solutions of Barcelo and Visser \cite{Barcelo:1999hq} is nevertheless called for. However, a quasinormal mode analysis is involved and will require a separate study on its own, as in the case of other wormholes (see e.g. \cite{Konoplya:2005et,Kim:2008zzj,Konoplya:2010kv,Konoplya:2016hmd,Bueno:2017hyj}).

While the study of the physical properties of these wormholes of Barcelo and Visser \cite{Barcelo:1999hq} represents a wide field, in particular, with respect to astrophysical applications, the most basic and important study to be performed is a study of the geodesics in these wormholes spacetimes. It is therefore the purpose of the present work to investigate the timelike, lightlike and spacelike geodesics and to discuss all possible types of orbits in these spacetimes, using effective potentials and parametric plots, and to present analytical solutions of the equations of motion. Since these are of elliptical types they can be expressed through the Weierstra\ss \ $\wp$-, $\sigma$- and $\zeta$-functions.

The paper is organized as follows. \autoref{JNWW:Sec} we give a brief description of the wormhole solutions and their properties. In \autoref{GeodEq:Sec} we provide the general set of the geodesic equations. We derive the analytical solutions in \autoref{Sol:Sec}, where we also discuss the effective potentials and classify the  possible orbit types. The orbits including the embeddings of their corresponding spacetimes are then illustrated in \autoref{Orbits:Sec}, while \autoref{Conclusion:Sec} gives our conclusions.

\section{Traversable wormholes supported by a massless conformally-coupled scalar field}
\label{JNWW:Sec}
Let us start by recalling the JNWW solution in the compact form found by Agnese and LaCamera \cite{Agnese:1985xj}

\begin{eqnarray}
\label{JNWW:Eq:dsm2}
&\resizebox{0.85\linewidth}{!}{$
ds_m^2 = -\left(1-\frac{2\eta}{r}\right)^{\cos\chi} dt^2 + \left(1-\frac{2\eta}{r}\right)^{-\cos\chi} dr^2 + \left(1-\frac{2\eta}{r}\right)^{1 - \cos\chi} r^2 (d\vartheta^2 + \sin^2\vartheta d\Phi^2)\;,$} \\[2ex]
&\phi_m = \sqrt{\frac{\kappa}{2}} \sin\chi \ln\left(1-\frac{2\eta}{r}\right)\,, 
\end{eqnarray}

\noindent with the scalar field $\phi_m$, and parameters $\chi$ and $\eta$. \\

Barcelo and Visser \cite{Barcelo:1999hq} reconsidered the JNWW solution in this form
when looking for static spherically symmetric solutions of the Einstein equations 

\begin{equation}
G_{\mu\nu} = \kappa T_{\mu\nu} \;,
\end{equation}

\noindent
where $\kappa = (8 \pi G_N)^{-1}$ and $G_{\mu\nu} = R_{\mu\nu} - \frac{1}{2} g_{\mu\nu} R$, with the new improved energy-momentum tensor $T_{\mu\nu}$ from \cite{Callan:1970ze}, 

\begin{eqnarray}
\centering
\label{Intro:Eq:EST1}
&T_{\mu \nu} = \nabla_\mu \phi_c \nabla_\nu \phi_c - \frac{1}{2} g_{\mu \nu} (\nabla \phi_c)^2 + \frac{1}{6} \left[G_{\mu\nu}\phi_c^2 - 2 \nabla_\mu(\phi_c \nabla_\nu \phi_c) + 2 g_{\mu\nu} \nabla^\lambda(\phi_c \nabla_\lambda \phi_c)\right], \\[2ex]
&\left(\square - \frac{1}{6} R\right)\phi_c = 0
\end{eqnarray}

\noindent with $\phi_c$ the conformally-coupled scalar field. (The improvement provided by this new energy-momentum tensor resides in the fact that its matrix elements are cutoff independent in the limit of large cutoff, whereas they are cutoff dependent in perturbation theory for most renormalizable field theories \cite{Callan:1970ze,Barcelo:1999hq}.)

The new improved energy-momentum tensor is traceless, thus also $R = 0$, yielding 

\begin{eqnarray}
\label{Intro:Eq:EST2}
&R_{\mu\nu} = \left(\kappa - \frac{1}{6} \phi_c^2\right)^{-1}\;\left(\frac{2}{3}\nabla_{\mu}\phi_c\nabla_{\nu}\phi_c - \frac{1}{6}g_{\mu\nu}(\nabla\phi_c)^2 - \frac{1}{3}\phi_c\nabla_{\mu}\nabla_{\nu} \phi_c\right), \\[3pt]
&\square\phi_c=0\;. 
\end{eqnarray}

Requiring that the metric $ds$ should  be conformal to the JNWW metric $ds_m$, i.e.,  $ds = \Omega(r) ds_m$ with conformal factor $\Omega(r)$, and should have vanishing scalar curvature, then leads to a second order differential  equation for $\Omega(r)$ with solutions \cite{Barcelo:1999hq}

\begin{equation}
\label{JNWW:Eq:ConfFac}
\Omega(r) = \alpha_+\; \left(1-\frac{2 \eta}{r}\right)^{\frac{\sin\chi}{2 \sqrt{3}}} + \alpha_-\; \left(1-\frac{2 \eta}{r}\right)^{-\frac{\sin\chi}{2 \sqrt{3}}}\,,
\end{equation}

\noindent where $\alpha_+$ and $\alpha_-$ are integration constants. The differential equation for the conformally coupled scalar field can then be integrated \cite{Barcelo:1999hq}. \\

With the parameter set $\lbrace\eta,\,\chi,\,\Delta\rbrace$, where $\Delta$ is an angle defined by

\begin{equation}
\tan\frac{\Delta}{2} = \frac{\alpha_+ - \alpha_-}{\alpha_+ + \alpha_-} = \frac{\bar {\alpha}_+ - 1}{\bar{\alpha}_+ + 1},\;\; \bar{\alpha}_+ = \frac{\alpha_+}{\alpha_-}\,
\label{baralpha}
\end{equation}

\noindent with range $\Delta \in (-\pi,\, \pi]$,  the whole set of solutions can be addressed. Depending on the choice of these parameters the properties of the metric can change dramatically. For example for $\chi = 0$ and arbitrary $\eta$ and $\Delta$, $ds^2$ yields the Schwarzschild metric. \\

To get traversable wormhole solutions, Barcelo and Visser found the appropriate parameter set to be $\lbrace\chi = \frac{\pi}{3},\, \Delta \notin \lbrace0,\,\frac{\pi}{2},\, \pi\rbrace\rbrace$. Introducing isotropic radial coordinates via $r = \bar{r} \left(1 + \frac{\eta}{2 \bar{r}}\right)^2$, the metric transforms into

\begin{equation}
\label{JNWW:Eq:Metrik}
ds^2 = \resizebox{0.75 \linewidth}{!}{$\left[\alpha_{+} \left(\frac{1-\frac{\eta}{2 \bar{r}}}{1+\frac{\eta}{2 \bar{r}}}\right)+\alpha_{-}\right]^2 \left[-dt^2+\left(1+\frac{\eta}{2 \bar{r}}\right)^4 \left[d\bar{r}^2+\bar{r}^2(d\vartheta^2+\sin^2\vartheta ~d\phi^2) \right]\right]$}\,,
\end{equation}

\noindent with the range $\bar{r} \in [0, \infty]$, where the radial location of the wormhole throat is given by

\begin{equation}
\label{JNWW:Eq:Throat}
\bar{r}_T = \frac{\eta}{2} \sqrt{\left|\frac{\bar{\alpha}_+ - 1}{\bar{\alpha}_+ + 1}\right|}\,,
\end{equation}

\noindent and the corresponding conformally-coupled scalar field is

\begin{equation}
\phi_c = \pm \sqrt{6 \kappa} ~\frac{\bar{\alpha}_+ \left(1 - \frac{\eta}{2 \bar{r}}\right) - \left(1 + \frac{\eta}{2 \bar{r}}\right)}{\bar{\alpha}_+ \left(1 - \frac{\eta}{2 \bar{r}}\right) + \left(1 + \frac{\eta}{2 \bar{r}}\right)}\,,
\end{equation}

\noindent which is a monotically increasing or decreasing function between the two asymptotically flat regions. \\

As pointed out by Barcelo and Visser, this monotonic behavior has physical consequences. In particular, the effective gravitational coupling constant

\begin{equation}
G_{\rm eff} = 8 \pi \left(\kappa - \frac{1}{6} \phi_c^2\right)^{-1}
\end{equation}

will be positive in one asymptotically flat region and negative in the other asymptotically flat region, and, with respect to the asymptotically flat region with positive effective gravitational coupling constant, the wormhole throat will be located in the region, where the effective gravitational coupling constant has changed sign \cite{Barcelo:1999hq}. Thus the change of sign does not happen at the throat [see also the discussion after \autoref{Sol:drdphi:Class:Eq:P1}]. In our study of the geodesics we will see that the geodesics are completely smooth when this sign change with diverging $G_{\rm eff}$ occurs. \\

Barcelo and Visser point out that the conformal coupling of the JNWW solution to the new improved energy-momentum tensor is well defined over the whole range of $\bar{r}$. Nevertheless the original metric $ds_m^2$ changes its sign for $\bar{r} \in (0, \frac{\eta}{2})$, which the conformal factor $\Omega(\bar{r})$ compensates. Thus only $\bar{r} > \frac{\eta}{2}$ is strictly speaking conformally related to the JNWW solution \cite{Barcelo:1999hq}. \\

Since $\alpha_+$ and $\alpha_-$ appear naturally in the metric, in the following $\bar{\alpha}_+$ will be used as a parameter for the wormhole solutions instead of $\Delta$. For traversable wormholes the range of $\bar{\alpha}_+$ is limited to $\bar{\alpha}_+ \in (0, 1)$. Negative $\bar{\alpha}_+$  won't be discussed, because in this case the wormhole spacetime is simply inverted.

\section{The geodesic equations}
\label{GeodEq:Sec}
Since the metric of the above traversable wormholes is static and spherically symmetric, we need to consider only  the equatorial plane with $\vartheta = \frac{\pi}{2}$ in the following. With the metric $ds^2$ from \autoref{JNWW:Eq:Metrik} we then get

\begin{eqnarray}
2 \Lagr &=& g_{\mu \nu} \dot{x}^{\mu} \dot{x}^{\nu} = \zeta \notag \\
&=& \left[\alpha_{+} \left(\frac{1-\frac{\eta}{2 \bar{r}}}{1+\frac{\eta}{2 \bar{r}}}\right)+\alpha_{-}\right]^2 \left[-\dot{t}^2+\left(1+\frac{\eta}{2 \bar{r}}\right)^4 \left[\dot{\bar{r}}^2+\bar{r}^2 ~\dot{\phi}^2\right]\right]\,, \label{GeodEq:Eq:Lagr}
\end{eqnarray}

\noindent
yielding timelike ($\zeta = -1$), lightlike ($\zeta = 0$) and spacelike ($\zeta = 1$) geodesics. We here consider spacelike geodesics for completeness. Note, however, that spacelike geodesics have also been studied, for instance, for Schwarzschild black hole spacetimes, where they allow to pass from one asymptotic region to the other (analytically extended) one \cite{Honig:1974br}.

The angular momentum $L$ and the energy  $\epsilon$ of the test particle are conserved,

\begin{equation}
L := \frac{\partial \Lagr}{\partial \dot{\phi}},\,\,-\epsilon := \frac{\partial \Lagr}{\partial \dot{t}} \; .
\end{equation}

\noindent
With these constants of motion the equations of motion in the equatorial plane become

\begin{eqnarray}
\left(\frac{d\bar{r}}{d\lambda}\right)^2 &=& \frac{\epsilon^2 + \zeta \left[\alpha_+ \left(\frac{1 - \frac{\eta}{2 \bar{r}}}{{1 + \frac{\eta}{2 \bar{r}}}}\right) + \alpha_- \right]^2}{\left[\alpha_+ \left(\frac{1 - \frac{\eta}{2 \bar{r}}}{{1 + \frac{\eta}{2 \bar{r}}}}\right) + \alpha_- \right]^4 \left(1 +\frac{\eta}{2 \bar{r}}\right)^4} - \frac{L^2}{\left[\alpha_+ \left(\frac{1 - \frac{\eta}{2 \bar{r}}}{{1 + \frac{\eta}{2 \bar{r}}}}\right) + \alpha_- \right]^4 \left(1 +\frac{\eta}{2 \bar{r}}\right)^8 \bar{r}^2} \label{GeoEq:Eq:drdl2} \\
\left(\frac{d\phi}{d\lambda}\right) &=& \frac{L}{\left[\alpha_+ \left(\frac{1 - \frac{\eta}{2 \bar{r}}}{{1 + \frac{\eta}{2 \bar{r}}}}\right) + \alpha_- \right]^2 \left(1 +\frac{\eta}{2 \bar{r}}\right)^4 \bar{r}^2} \label{GeoEq:Eq:dphidl} \\
\left(\frac{dt}{d\lambda}\right) &=& \frac{\epsilon}{\left[\alpha_+ \left(\frac{1 - \frac{\eta}{2 \bar{r}}}{{1 + \frac{\eta}{2 \bar{r}}}}\right) + \alpha_- \right]^2} \label{GeoEq:Eq:dtdl}
\end{eqnarray}

\section{Solutions of the geodesic equations}
\label{Sol:Sec}
Whereas the equations of motion (\ref{GeoEq:Eq:drdl2}), (\ref{GeoEq:Eq:dphidl}) and (\ref{GeoEq:Eq:dtdl}) cannot be solved analytically by simple means, the equations for $(d\bar{r}/d\phi)$ and $(d\bar{r}/dt)$ 

\begin{eqnarray}
\left(\frac{d\bar{r}}{d\phi}\right)^2 &=& \frac{\left(\epsilon^2 + \zeta \left[\alpha_{+} \left(\frac{1-\frac{\eta}{2 \bar{r}}}{1+\frac{\eta}{2 \bar{r}}}\right) + \alpha_{-} \right]^2 \right) \left(1 +\frac{\eta}{2 \bar{r}}\right)^4\bar{r}^4 - L^2 \bar{r}^2}{L^2} \label{Sol:Eq:drdphi2} \\
\left(\frac{d\bar{r}}{dt}\right)^2 &=& \frac{\epsilon^2 + \zeta \left[\alpha_+ \left(\frac{1 - \frac{\eta}{2 \bar{r}}}{{1 + \frac{\eta}{2 \bar{r}}}}\right) + \alpha_- \right]^2}{\epsilon^2 \left(1 +\frac{\eta}{2 \bar{r}}\right)^4} - \frac{L^2}{\epsilon^2 \left(1 +\frac{\eta}{2 \bar{r}}\right)^8 \bar{r}^2} \label{Sol:Eq:drdt2}
\end{eqnarray}

\noindent
can be solved in terms of Weierstra\ss \ functions. In \hyperref[Sol:drdphi:Sec]{Sec. IV A} we derive the solution for the $(d\bar{r}/d\phi)$ motion. We classify the possible orbit types in \hyperref[Sol:drdphi:Class:Sec]{Sec. IV A.1}, and we derive the solution for the $(d\bar{r}/dt)$ motion in \hyperref[Sol:drdt:Sec]{Sec. IV B}.

\subsection{Solution for the \boldmath{$(d\bar{r} / d\phi)$}-motion}
\label{Sol:drdphi:Sec}
We first substitute in \autoref{Sol:Eq:drdphi2} the radial coordinate $\bar{r} = \eta (x - \frac{1}{2})$, where the new radial coordinate $x$ has a range of $x \in \left[\frac{1}{2},\, \infty\right]$. The radial coordinate of the throat is then given by $x_T = \frac{1}{2} (\sqrt{|(\bar{\alpha}_+ - 1)/(\bar{\alpha}_+ + 1)|} + 1)$. This yields an equation which can be written in terms of a polynomial of fourth order

\begin{equation}
\left(\frac{dx}{d\phi}\right)^2 = a_4 x^4 + a_3 x^3 + a_2 x^2 + a_1 x + a_0 
\end{equation}

\noindent with the coefficients

\begin{eqnarray}
a_4 &=& \bar{\epsilon}^2 \zeta (\bar{\alpha}_+ + 1)^2 / \tilde{L}^2\,, \notag \\
a_3 &=& 2 \zeta ({\bar{\alpha}_+}^2 + \bar{\alpha}_+) / \tilde{L}^2\,, \notag \\
a_2 &=& \zeta {\bar{\alpha}_+}^2 / \tilde{L}^2 - 1\,, \\
a_1 &=& 1\,, \notag \\
a_0 &=& 1 / 4\,. \notag
\end{eqnarray}

With another substitution $x = \pm \frac{1}{u} + x_1$, where $x_1$ is a zero of $(dx/d\phi)^2$, we obtain a polynomial of third order

\begin{equation}
\left(\frac{du}{d\phi}\right)^2 = b_3 u^3 + b_2 u^2 + b_1 u + b_0\,.
\end{equation}

\noindent
The additional substitution $u = \frac{1}{b_3} (4 y - \frac{b_2}{3})$ then yields the Weierstra\ss \ form

\begin{equation}
\label{Sol:drdphi:Eq:dydphi}
\left(\frac{dy}{d\phi}\right)^2 = 4 y^3 - g_2 y - g_3\,,
\end{equation}

\noindent where

\begin{eqnarray}
g_2 &=& \frac{1}{16} \left(\frac{4}{3}\;b_2^2 - 4 b_1 b_3\right)\,, \notag \\
g_3 &=& \frac{1}{16} \left(\frac{1}{3}\; b_1 b_2 b_3 - \frac{2}{27} b_2^3 - b_0 b_3^2\right)\,.
\end{eqnarray}

Then $y(\phi) = \wp(\phi - c_\text{in}, g_2, g_3)$ solves \autoref{Sol:drdphi:Eq:dydphi}, where $c_\text{in} = \phi_0 + \int_{y_0}^\infty \frac{dz}{\sqrt{4 z^3 - g_2 z - g_3}}$ and $y_0 = \pm\frac{b_3}{4(x_0 - x_1)} - \frac{b_2}{12}$. Thus \autoref{Sol:Eq:drdphi2} has the solution

\begin{equation}
\label{Sol:drdphi:Eq:xsol}
x(\phi) = \pm \frac{b_3}{4 \wp(\phi - c_\text{in},\, g_2,\, g_3) - \frac{b_2}{3}} + x_1\,.
\end{equation}

\subsubsection{Classification of the \boldmath{$(d\bar{r} / d\phi)$}-motion}
\label{Sol:drdphi:Class:Sec}
In the following we will show that only a small set of orbit types are found when the  $(d\bar{r} / d\phi)$-motion is investigated. Those orbit types for timelike, lightlike and spacelike geodesics consist of the following possible orbits

\begin{itemize}
\item \textit{Transit orbit} (TO): A geodesic coming from one asymptotic region approaches the throat of the wormhole, crosses it and reaches the other asymptotic region, thus traversing the whole wormhole spacetime.
\item \textit{Bound orbit} (BO): A geodesic moves continuously around the wormhole, remaining on the same side of the throat. 
\item \textit{Escape orbit} (EO): A geodesic coming from one asymptotic region approaches the throatof the wormhole, but instead of crossing the throat it reaches a turning point and returns to the asymptotic region, where it came from.
\item \textit{Two world escape orbit} (TWEO): Similar to the EO with the difference that the geodesic crosses the throat before it reaches its turning point, from where it returns to the asymptotic region, where it came from, passing the throat on its way a second time.
\item \textit{Unstable circular orbit} (UCO): A bound orbit which moves around the wormhole on a circular path. These orbits are highly unstable due  to the fact that their energy is equal to the local potential maximum.
\end{itemize}

Clearly, geodesic motion is only possible if $(d\bar{r}/d\phi)^2 \geqslant 0$. This leads to the following condition for the energy $\epsilon$ of a particle for physically allowed motion,

\begin{equation}
\label{Sol:drdphi:Class:Eq:pre_epsilon}
\epsilon^2 \geqslant \frac{L^2}{\left(1+\frac{\eta}{2 \bar{r}}\right)^4 \bar{r}^2} - \zeta \left[\alpha_{+} \left(\frac{1-\frac{\eta}{2 \bar{r}}}{1+\frac{\eta}{2 \bar{r}}}\right)+\alpha_{-}\right]^2 \;.
\end{equation}

\noindent Note, that the right hand side may be considered as an effective potential, consisting of a centrifugal part (the first term) and a gravitational part (the second term).

We now perform two coordinate transformations to simplify \autoref{Sol:drdphi:Class:Eq:pre_epsilon}. We first introduce again the radial coordinate  $x = \frac{\bar{r}}{\eta} + \frac{1}{2}$, as in the solution for $(d\bar{r}/d\phi)$, and then we transform to the radial coordinate $w = \frac{1}{2 x}$. This new radial coordinate has the range $w \in [0, 1]$, and the throat is located at $w_T = (\sqrt{|(\bar{\alpha}_+ - 1)/(\bar{\alpha}_+ + 1)|} + 1)^{-1}$. We further employ the scaled quantities $\bar{\alpha}_+$ (\autoref{baralpha}), $\tilde{L}=L/(\eta \alpha_-)$ and $\bar{\epsilon}=\epsilon/\alpha_-$ to obtain the condition

\begin{equation}
\label{Sol:drdphi:Class:Eq:epsilon}
\bar{\epsilon}^2 \geqslant \underbrace{4 \tilde{L}^2 w^2 (1-w)^2 - \zeta \left[\bar{\alpha}_{+} (2 w-1) - 1\right]^2}_{:=P(\bar{\alpha}_{+}, \tilde{L}, w, \zeta)}
\end{equation}

Here $P(\bar{\alpha}_{+}, \tilde{L}, w, \zeta)$ is a polynomial of fourth order, which serves as an effective potential and thus helps characterizing the possible orbits. In the two asymptotic regions with radial coordinates $w=0$ and $w=1$, the polynomial assumes the values

\begin{eqnarray}
P(\bar{\alpha}_{+}, \tilde{L}, w=0, \zeta) := P_0(\bar{\alpha}_+,\; \zeta) = -\zeta (\bar{\alpha}_+ + 1)^2\,, \label{Sol:drdphi:Class:Eq:P0} \\
P(\bar{\alpha}_{+}, \tilde{L}, w=1, \zeta) := P_1(\bar{\alpha}_+,\; \zeta) = -\zeta (\bar{\alpha}_+ - 1)^2\,. \label{Sol:drdphi:Class:Eq:P1}
\end{eqnarray}

\noindent 
These limiting values are determined only by the gravitational part of the effective  potential, since the centrifugal part vanishes asymptotically.

We note that the gravitational part is monotonically decreasing (increasing) in the interval $w \in [0, 1]$ for timelike (spacelike) geodesics, while the centrifugal part is symmetric with respect to $w=1/2$, where it reaches its maximum. Expressed in terms of the radial coordinate $\bar r$ this value corresponds to $\bar r=\eta/2$. At precisely this value the conformal scalar field assumes the values $\phi_c=\pm \sqrt{6 \kappa}$; i.e., the factor $\left(\kappa - \frac{1}{6} \phi_c^2\right)^{-1}$ on the right hand side of the Einstein equations (\ref{Intro:Eq:EST2}), and thus the effective Newton constant $G_{\rm eff}$, diverges. We also note that because of our choice of orientation of the wormhole ($\bar{\alpha}_+ > 0$), the throat is always located in the rear part of the physical $w$-interval, i.e., $w_T \in (0.5,1)$.

To classify the possible orbits, we next need to consider the minima and maxima of the polynomial $P$. We obtain them by calculating the derivative $P' = \frac{\partial P}{\partial w}$ and applying Cardanos' method to solve $P' = 0$ in order to get its zeros $w_1$, $w_2$ and $w_3$. From the discriminant $\Delta_\text{dis}(\tilde{L}, \bar{\alpha}_+, \zeta)$ of $P'$ we learn for which parameters the zeros of the polynomial are real. The discriminant is given by

\begin{equation}
\label{Sol:drdphi:Class:Eq:dis}
\Delta_\text{dis}(\tilde{L}, \bar{\alpha}_+, \zeta) = \frac{1}{1728 \tilde{L}^6} (-\tilde{L}^6 - 6 \tilde{L}^4 \bar{\alpha}_{+}^2 \zeta + (-12 \bar{\alpha}_{+}^4 + 27 \bar{\alpha}_{+}^2)\; \tilde{L}^2 \zeta^2 - 8 \tilde{\alpha}_{+}^6 \zeta^3)\;.
\end{equation}

\noindent When $\Delta_\text{dis} < 0$, all zeros of $P'$ are real. For fixed $\bar{\alpha}_+$ and $\zeta$ this is the case, when the angular momentum exceeds a critical value, $\tilde{L} > \tilde{L}_\text{crit}$. Then the three zeros of the derivative $P'$ are given by

\begin{eqnarray}
w_1 &=& \sqrt{-\frac{4}{3}p}~\cos\left(\frac{1}{3}\arccos\left(-\frac{q}{2}\sqrt{-\frac{27}{p^3}}\;\right)\right)+\frac{1}{2}\,, \\
w_{2,3} &=& -\sqrt{-\frac{4}{3}p}~\cos\left(\frac{1}{3}\arccos\left(-\frac{q}{2}\sqrt{-\frac{27}{p^3}}\;\right)\pm\frac{\pi}{3}\right)+\frac{1}{2} \,,
\end{eqnarray}

\noindent with $p = -\frac{1}{2}~\frac{\bar{\alpha}_+^2\;\zeta}{\tilde{L}^2}-\frac{1}{4}$ and $q = \frac{1}{4}~ \frac{\bar{\alpha}_+\; \zeta}{\tilde{L}^2}$. Let us denote the maximum and the minima of the polynomial as follows

\begin{eqnarray}
\label{Sol:drdphi:Class:Eq:Pmaxmin}
P_{\text{max}}(\bar{\alpha}_+, \tilde{L}, \zeta) &:=& P(\bar{\alpha}_+, \tilde{L}, w=w_2, \zeta)\,, \notag \\
P_{\text{min}}(\bar{\alpha}_+, \tilde{L}, \zeta) &:=& \begin{cases}
P(\bar{\alpha}_+, \tilde{L}, w=w_1, \zeta), & \zeta = -1\;, \\
P(\bar{\alpha}_+, \tilde{L}, w=w_3, \zeta), & \zeta = 1\;.
\end{cases} \notag
\end{eqnarray}

\noindent Concerning $w_1$ and $w_3$ we note that only one of them is located in the allowed range of $w$, depending on the geodesic type. \\

\begin{figure}
\centering
\subfloat[$\zeta = -1$]{\label{Sol:drdphi:Class:Fig:lcrit:z-1}\includegraphics[width=0.49\linewidth]{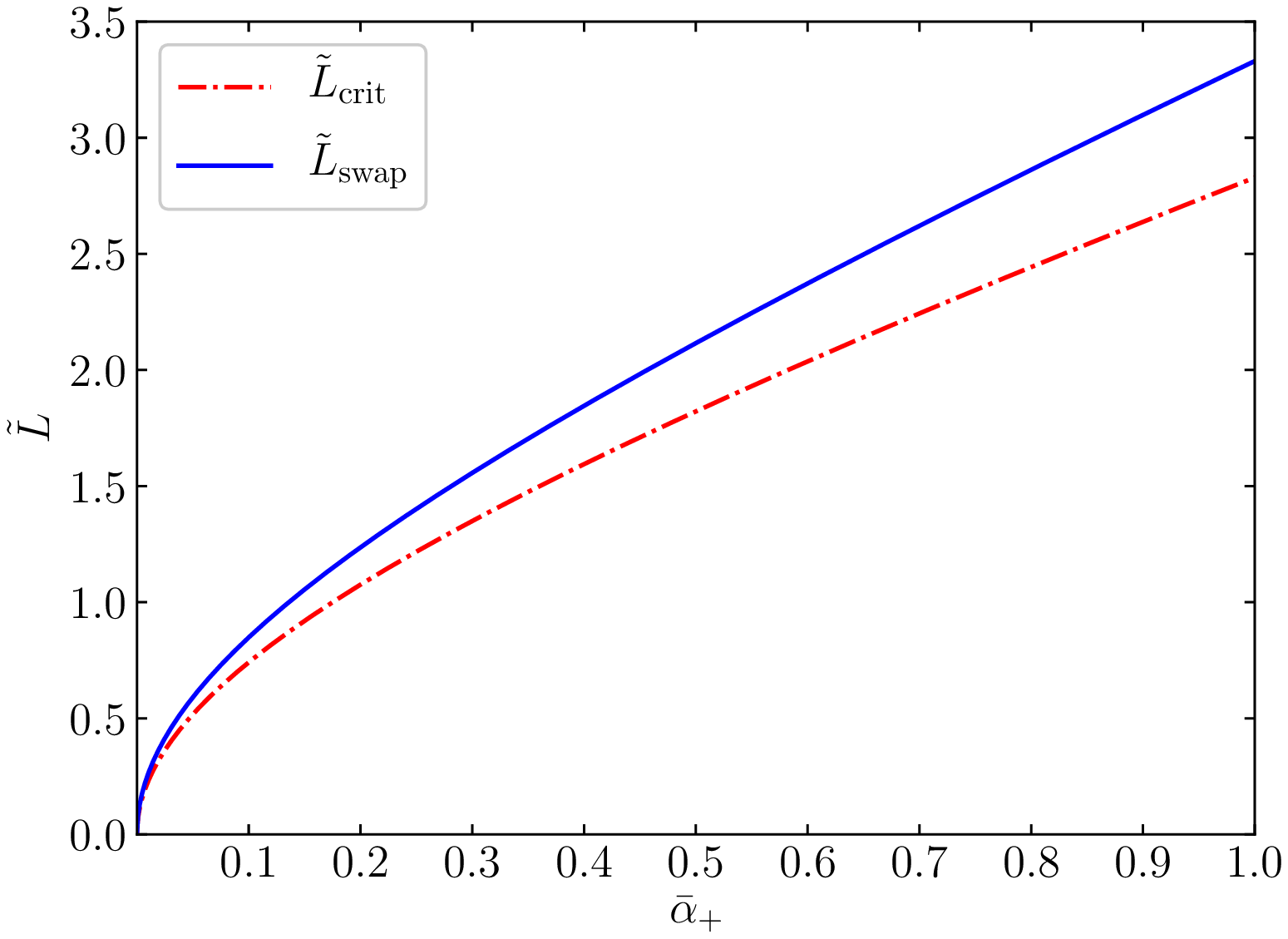}}
\subfloat[$\zeta = 1$]{\label{Sol:drdphi:Class:Fig:lcrit:z1}\includegraphics[width=0.49\linewidth]{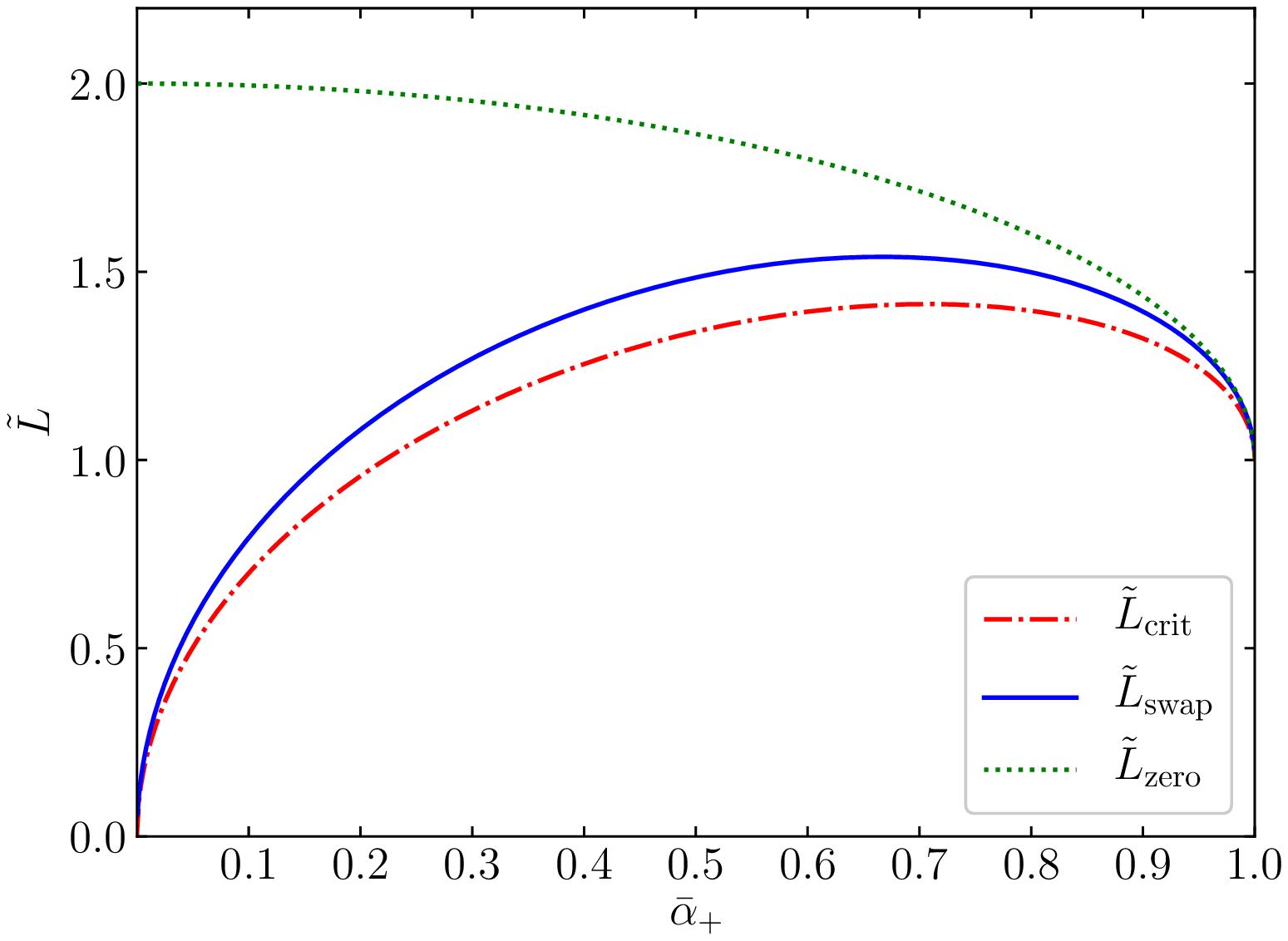}}
\caption{Dependence of the characteristic angular momenta $\tilde{L}_\text{crit}$ and $\tilde{L}_\text{swap}$ on the wormhole parameter $\bar{\alpha}_+$ for timelike geodesics (a) and spacelike geodesics (b). The latter includes $\tilde{L}_\text{zero}$. For timelike geodesics the gap between $\tilde{L}_\text{crit}$ and $\tilde{L}_\text{swap}$ increases monotonically with $\bar{\alpha}_+$. For spacelike geodesics only $\tilde{L}_\text{zero}$ is relevant, since the energy of particles should be a real number.}
\label{Sol:drdphi:Class:Fig:lcrit}
\end{figure}

Since the effective potential is a superposition of the monotonic gravitational part and the symmetric centrifugal part, the orbit types for timelike and spacelike geodesics change at certain characteristic values of the angular momentum. Besides the critical angular momentum $\tilde{L}_{\rm crit}$, where a stationary inflection point of the polynomial $P$ arises, there are two more such characteristic values. These are the angular momentum $\tilde{L}_\text{swap}$, where $P_{\text{max}}(\bar{\alpha}_+, \tilde{L}, \zeta)$ becomes larger than $P_0(\bar{\alpha}_+,\; \zeta)$, and (only for spacelike geodesics) the angular momentum $\tilde{L}_\text{zero}$, beyond which the polynomial $P(\bar{\alpha}_{+}, \tilde{L}, w, \zeta)$ can assume positive values. The dependence of these characteristic values of the angular momentum on $\bar{\alpha}_+$ is shown in \autoref{Sol:drdphi:Class:Fig:lcrit}. \\

\begin{figure}
\centering
\subfloat[$\tilde{L} < \tilde{L}_\text{crit}$]{\label{Sol:drdphi:Class:Fig:PolyPz-1:no3}\includegraphics[width=0.44\linewidth]{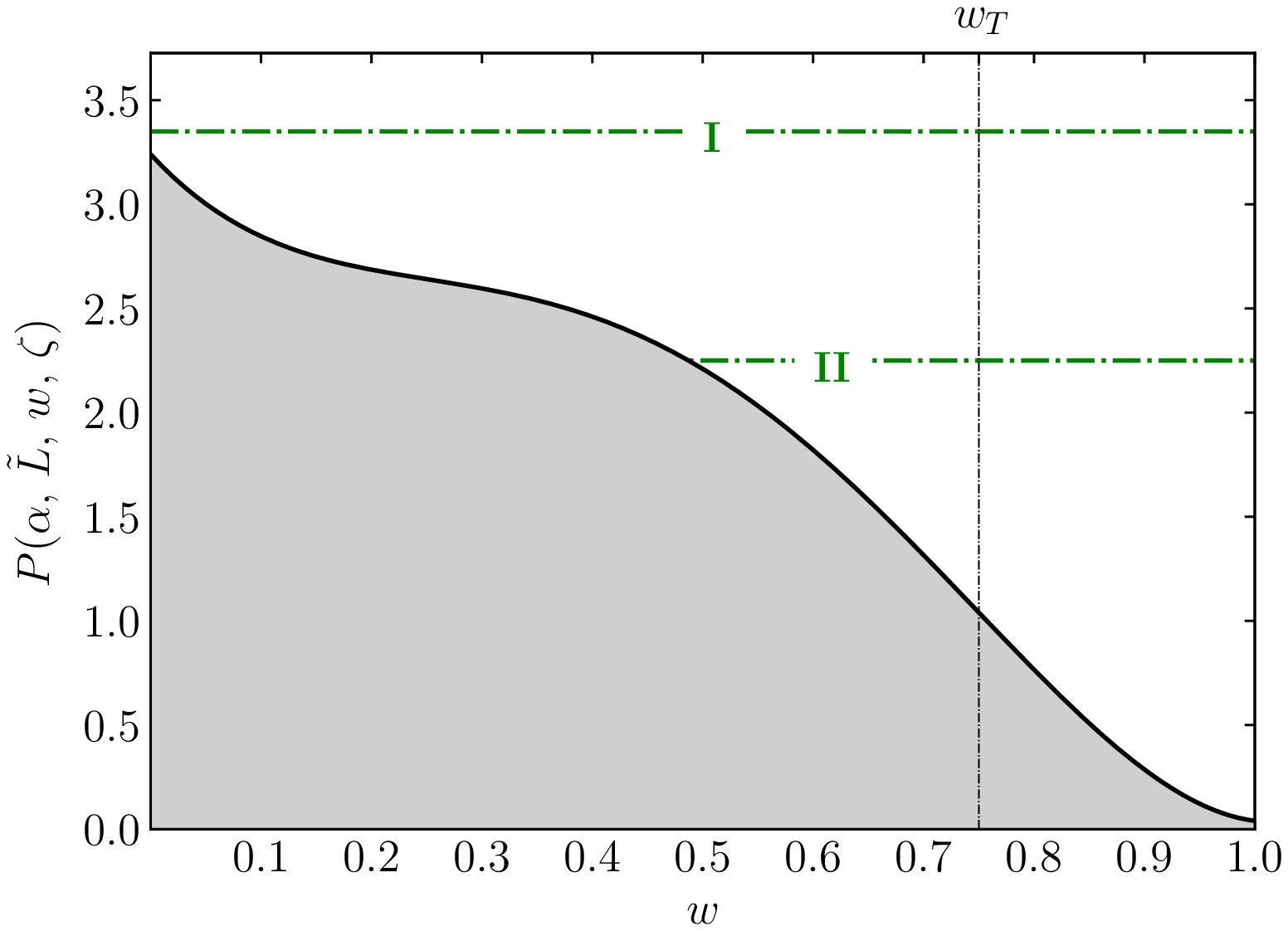}}
\subfloat[$\tilde{L} = \tilde{L}_\text{crit}$]{\label{Sol:drdphi:Class:Fig:PolyPz-1:L=Lcrit}\includegraphics[width=0.44\linewidth]{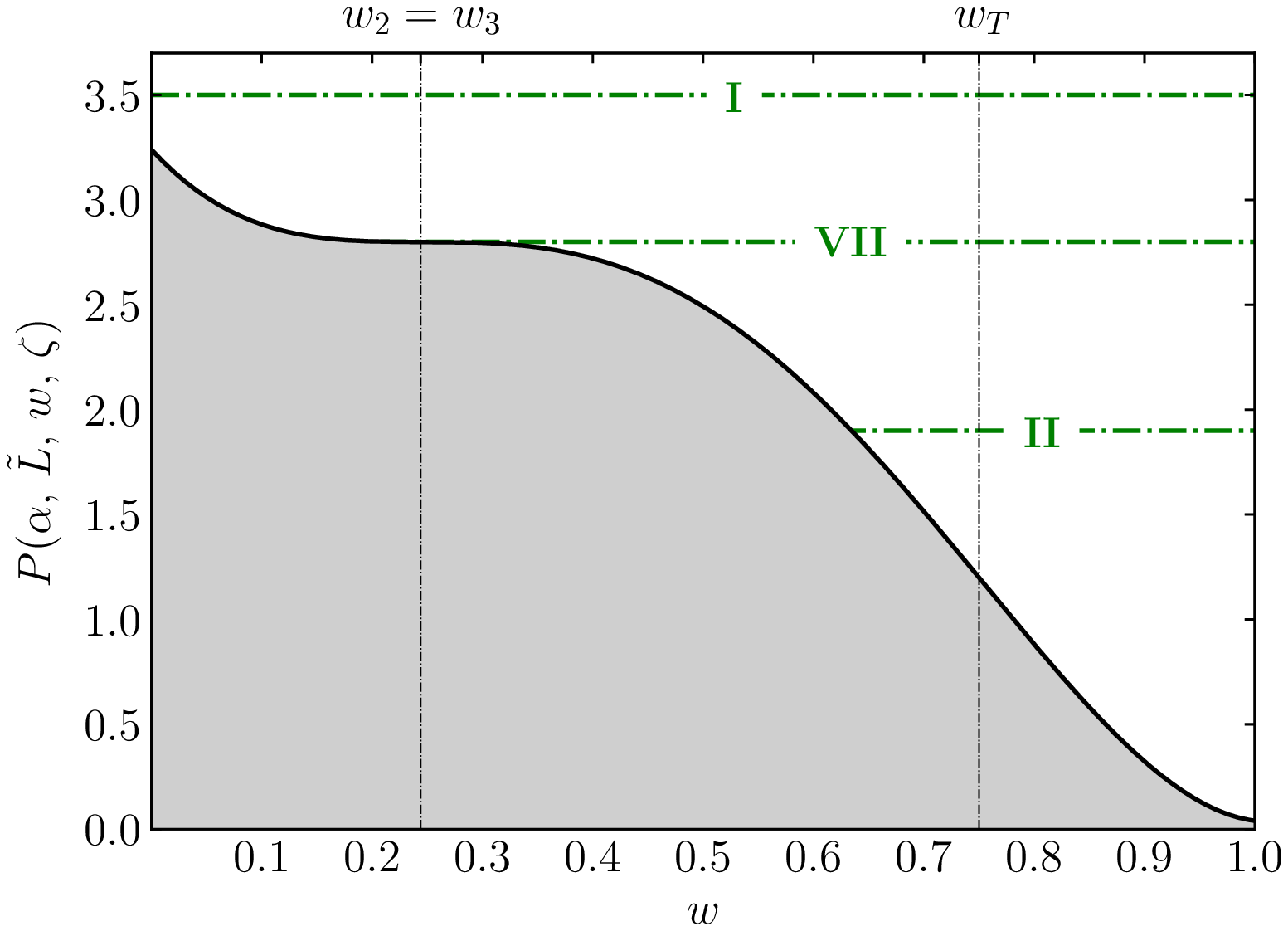}} \\
\subfloat[$\tilde{L}_\text{crit} < \tilde{L} < \tilde{L}_\text{swap}$]{\label{Sol:drdphi:Class:Fig:PolyPz-1:Lcrit}\includegraphics[width=0.44\linewidth]{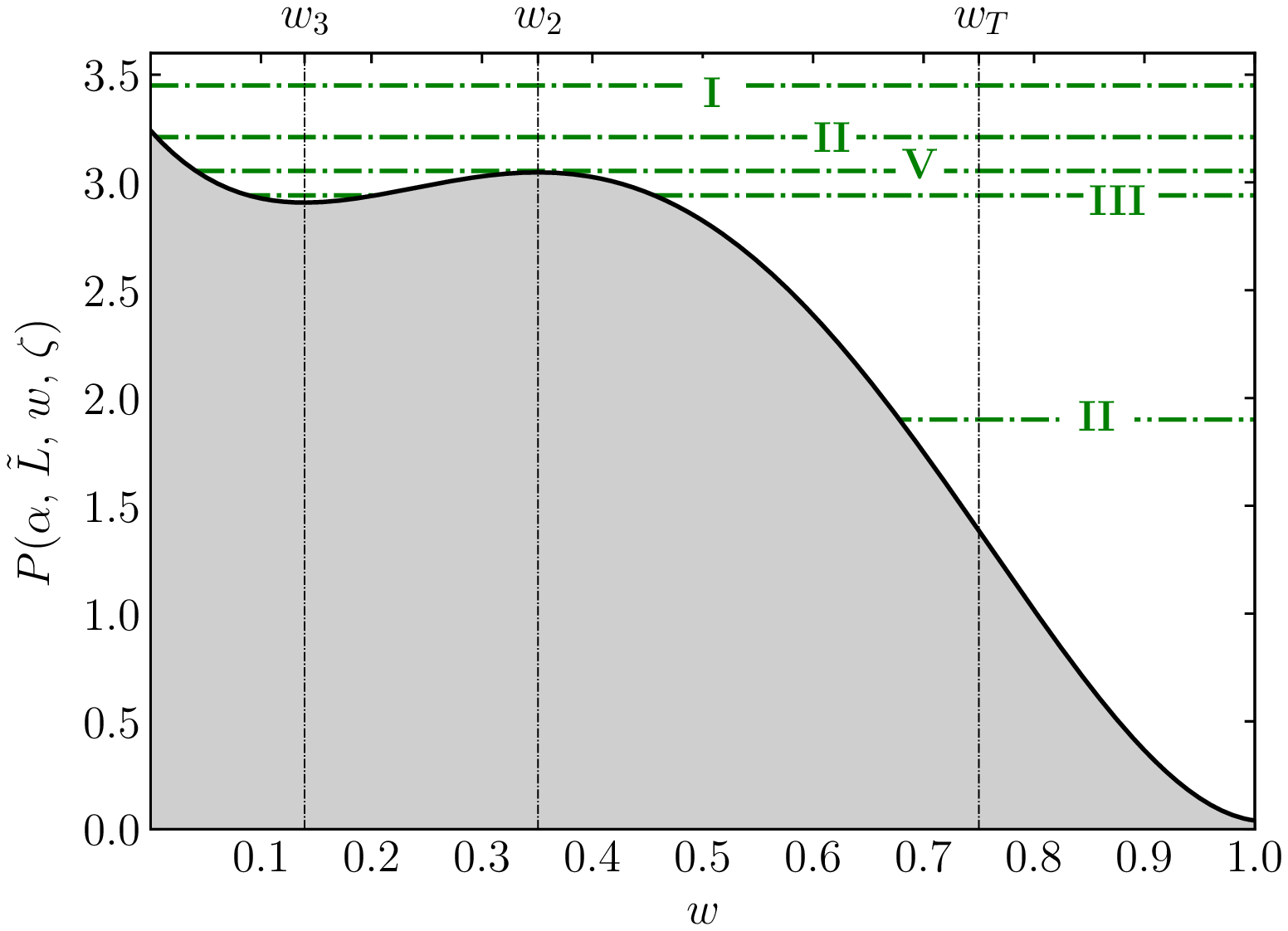}}
\subfloat[$\tilde{L}_\text{swap} < \tilde{L}$]{\label{Sol:drdphi:Class:Fig:PolyPz-1:Lswap}\includegraphics[width=0.44\linewidth]{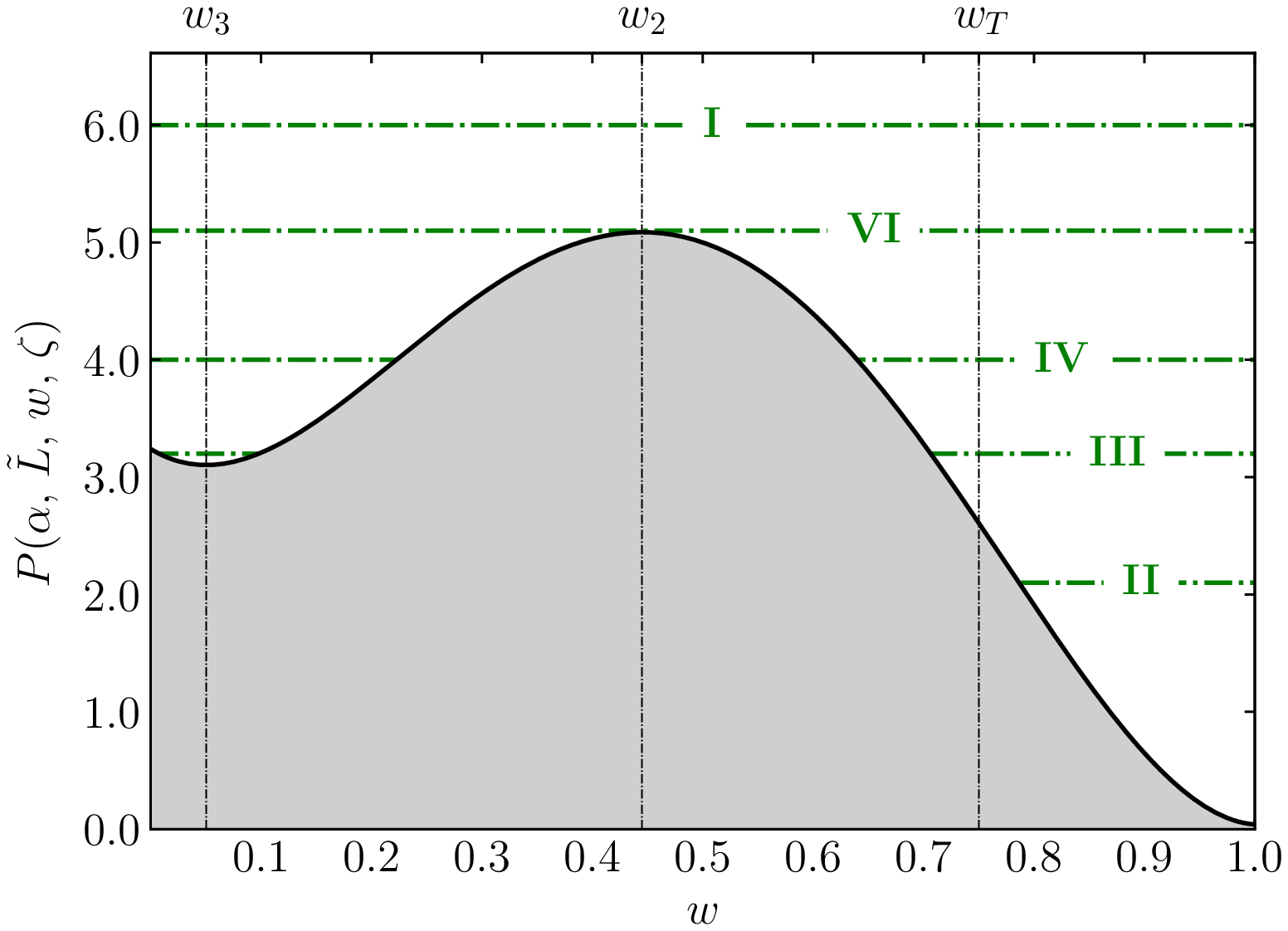}}
\caption{Polynomial $P(\bar{\alpha}_{+}, \tilde{L}, w, \zeta)$ for timelike geodesics and different characteristic values of $\tilde{L}$ versus the radial coordinate $w$. The green dotted lines represent the different orbit types possible for the corresponding energies $\bar{\epsilon}^2$. The grey areas represent physically forbidden zones, where no motion can occur. The wormhole parameter for all plots is $\bar{\alpha} = 0.95$.}
\label{Sol:drdphi:Class:Fig:PolyPz-1}
\end{figure}

To clarify the classification of the orbit types, we illustrate the effective potential as described by the polynomial $P$ in \autoref{Sol:drdphi:Class:Fig:PolyPz-1} for several values of the angular momentum, which allows for different orbit types. For the timelike geodesics three different cases for the angular momentum $\tilde{L}$ arise and are shown in \autoref{Sol:drdphi:Class:Fig:PolyPz-1}. For $\tilde{L} < \tilde{L}_\text{crit}$ the polynomial is monotonic. Here only transit and escape orbits exist. For $\tilde{L} = \tilde{L}_\text{crit}$ the polynomial $P$ possesses a double zero in the allowed physical range of $w$. Here unstable circular, transit and escape orbits exist. For $\tilde{L}_\text{crit} < \tilde{L} < \tilde{L}_\text{swap}$ also bound orbits are possible due to the minimum of the polynomial $P$. An unstable circular orbit can occur as well due to the existence of a maximum. Finally, for $\tilde{L}_\text{swap} < \tilde{L}$ the maximum $P_{\text{max}}$ becomes larger than $P_0$, allowing for one additional orbit type (see Fig. \ref{Sol:drdphi:Class:Fig:PolyPz-1:Lswap}).

\begin{figure}[H]
\centering
\subfloat[timelike geodesics ($\zeta = -1$)]{\includegraphics[width=0.34\linewidth]{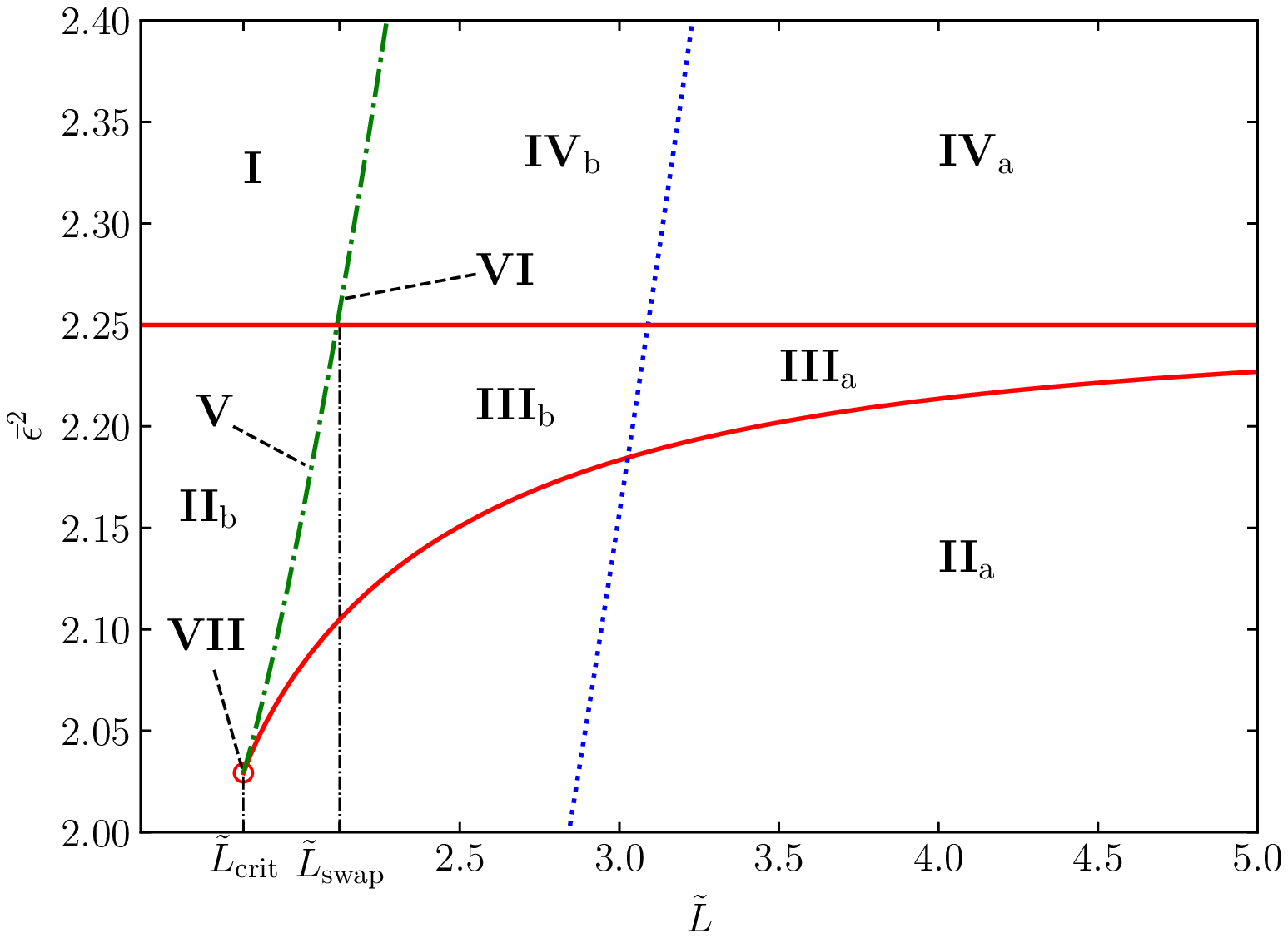}\label{Sol:drdphi:Class:Fig:ParamPlot:z-1}}
\subfloat[lightlike geodesics ($\zeta = 0$)]{\includegraphics[width=0.33\linewidth]{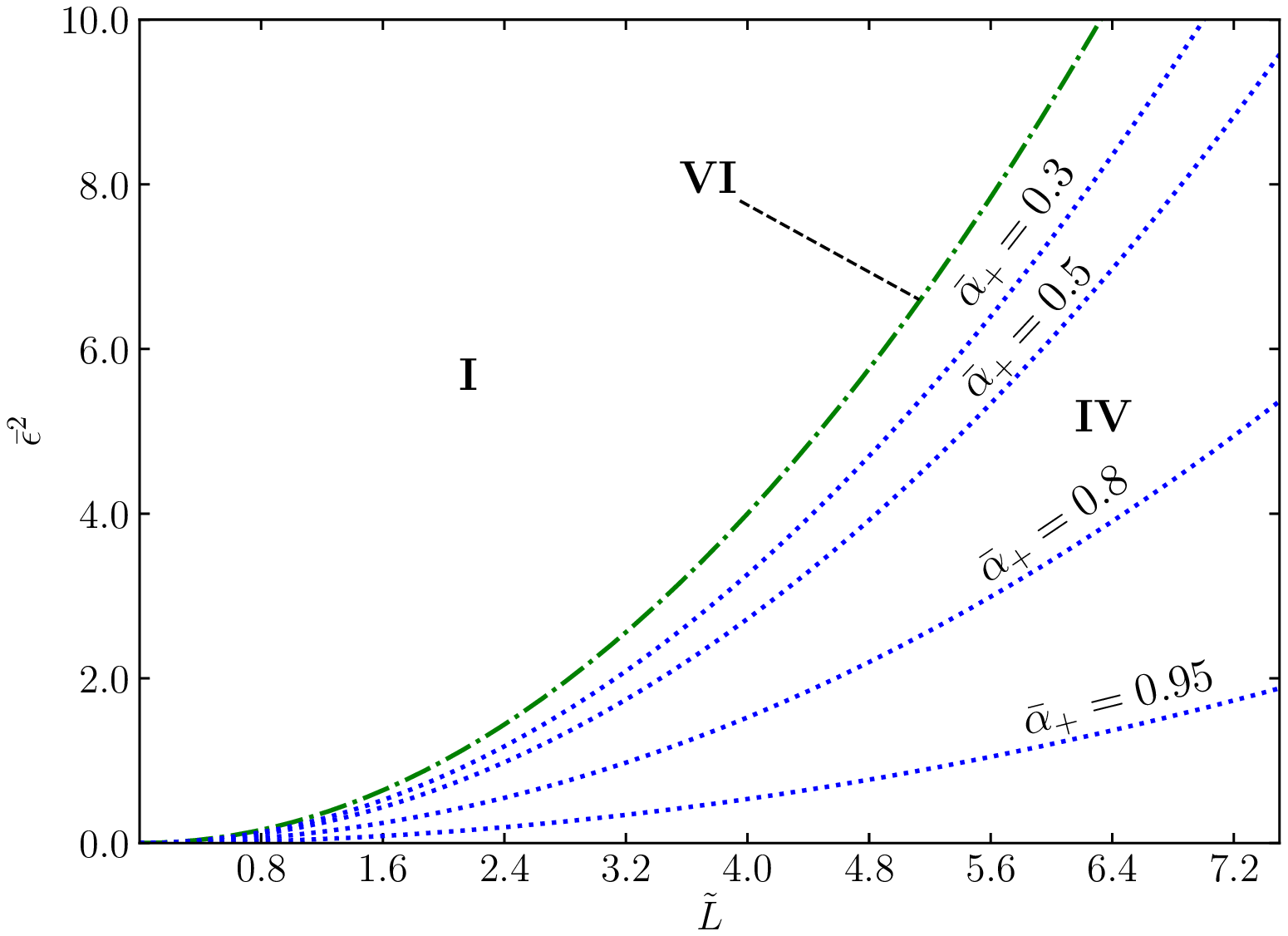}\label{Sol:drdphi:Class:Fig:ParamPlot:z0}}
\subfloat[spacelike geodesics ($\zeta = 1$)]{\includegraphics[width=0.33\linewidth]{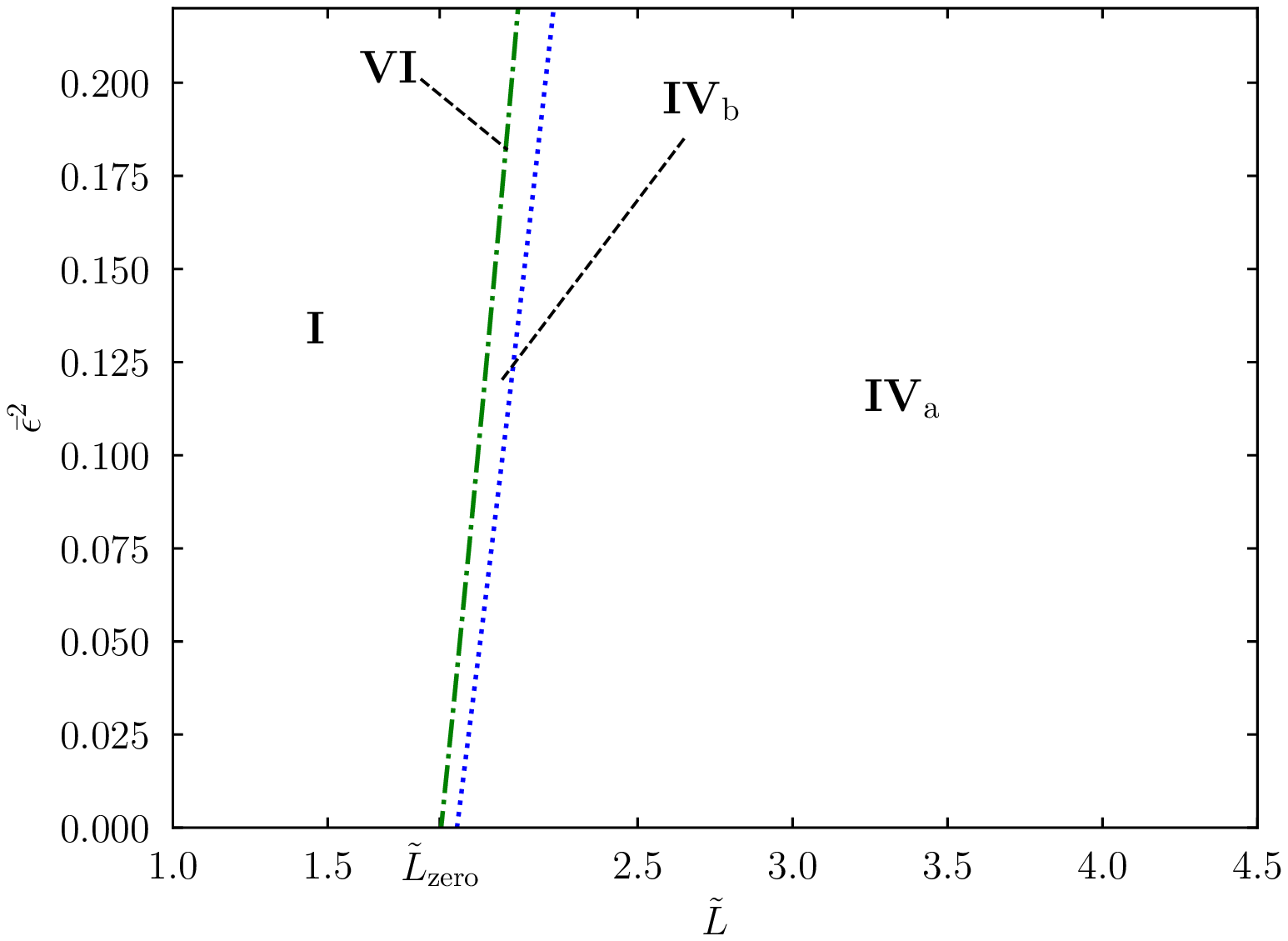}\label{Sol:drdphi:Class:Fig:ParamPlot:z1}}
\caption{Parametric plots for timelike (a), lightlike (b) and spacelike (c) geodesics with energy $\bar{\epsilon}^2$ versus angular momentum $\tilde{L}$. All parametric plots are evaluated for $\bar{\alpha}_+ = 0.50$. When $\bar{\alpha}_+$ is varied, the overall structure of the orbit regions with respect to each other is preserved, while the size of the regions can increase or decrease. The blue line (dotted) in the plots represents the wormhole barrier $P_T(\bar{\alpha}_{+}, \tilde{L}, \zeta)$. This curve also varies with $\bar{\alpha}_+$, as demonstrated for the lightlike geodesics. Spacelike geodesics are only shown for $\bar{\epsilon}^2 > 0$. Red lines (solid) limit the regions. The green line (dash-dotted) represents a region in itself as well as a limit between regions. For timelike geodesics the orbit type on the line changes at $\tilde{L}_\text{swap}$. The circle in \autoref{Sol:drdphi:Class:Fig:ParamPlot:z-1} indicates the point where the Type \textbf{\RN{7}} orbits occur.}
\label{Sol:drdphi:Class:Fig:ParamPlot}
\end{figure}

Turning to the lightlike orbits, we note that only the centrifugal part of the effective potential is present. Thus there is no dependence on $\bar{\alpha}_+$. The effective potential vanishes in the asymptotic regions $w=0$ and $w=1$, and has at $w=0.5$ its maximum, whose height depends only on the size of the angular momentum. Thus transit and escape orbits exist for lightlike geodesics as well as unstable circular orbits, illustrated in Fig. \ref{Sol:drdphi:Class:Fig:ParamPlot:z0} by a green line (dash-dotted), which form a photon sphere. Unstable circular orbits can occur for spacelike and timelike geodesics as well, when their energy corresponds to the local maximum.

\begin{figure}[H]
\centering
\subfloat[$\zeta = 0,\, \bar{\alpha}_+ = 0.85$]{\includegraphics[width=0.48\linewidth]{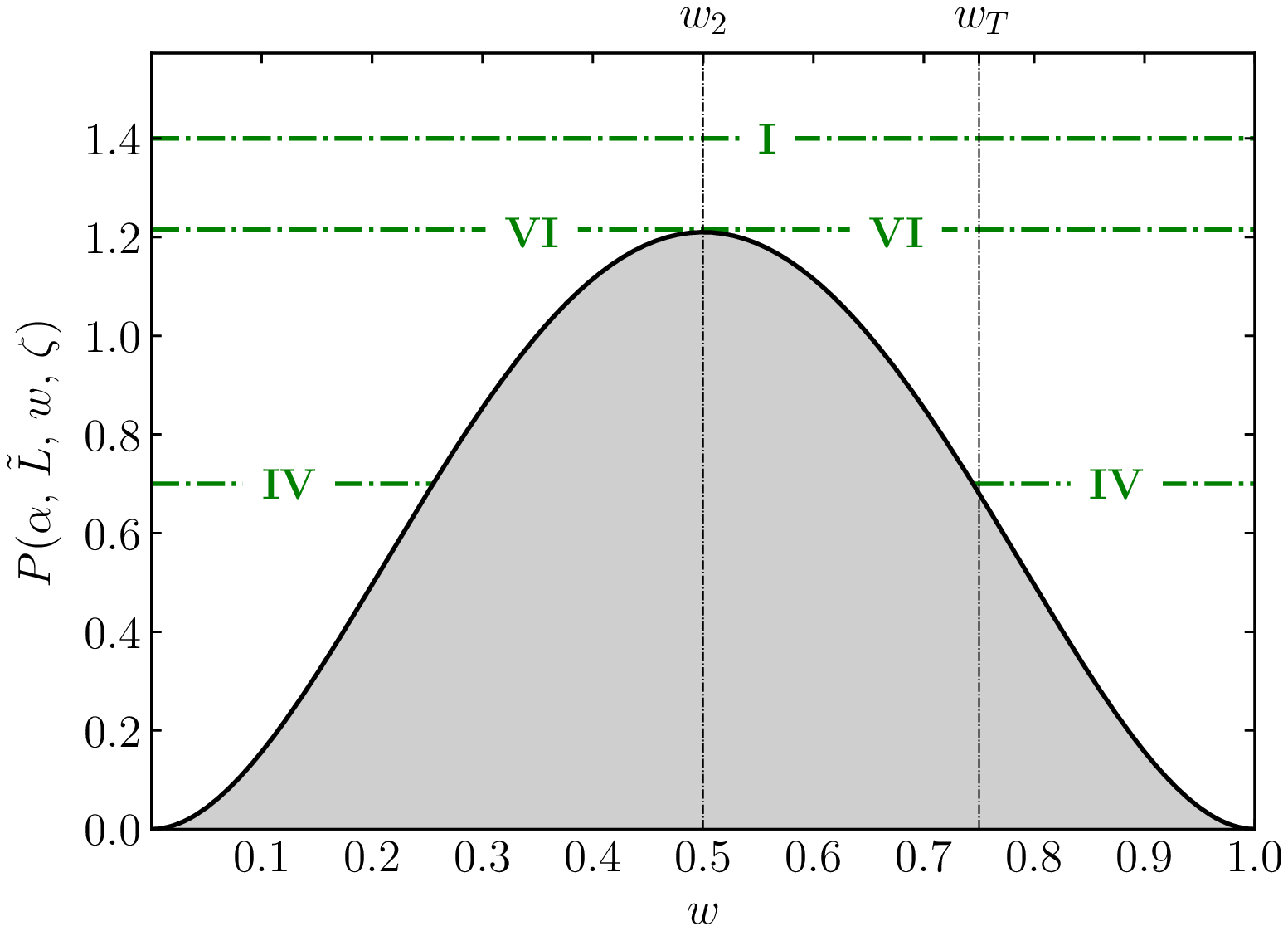}\label{Sol:drphi:Class:Fig:otherPoly:zeta0}}
\subfloat[$\zeta = 1,\, \bar{\alpha}_+ = 0.85$]{\includegraphics[width=0.48\linewidth]{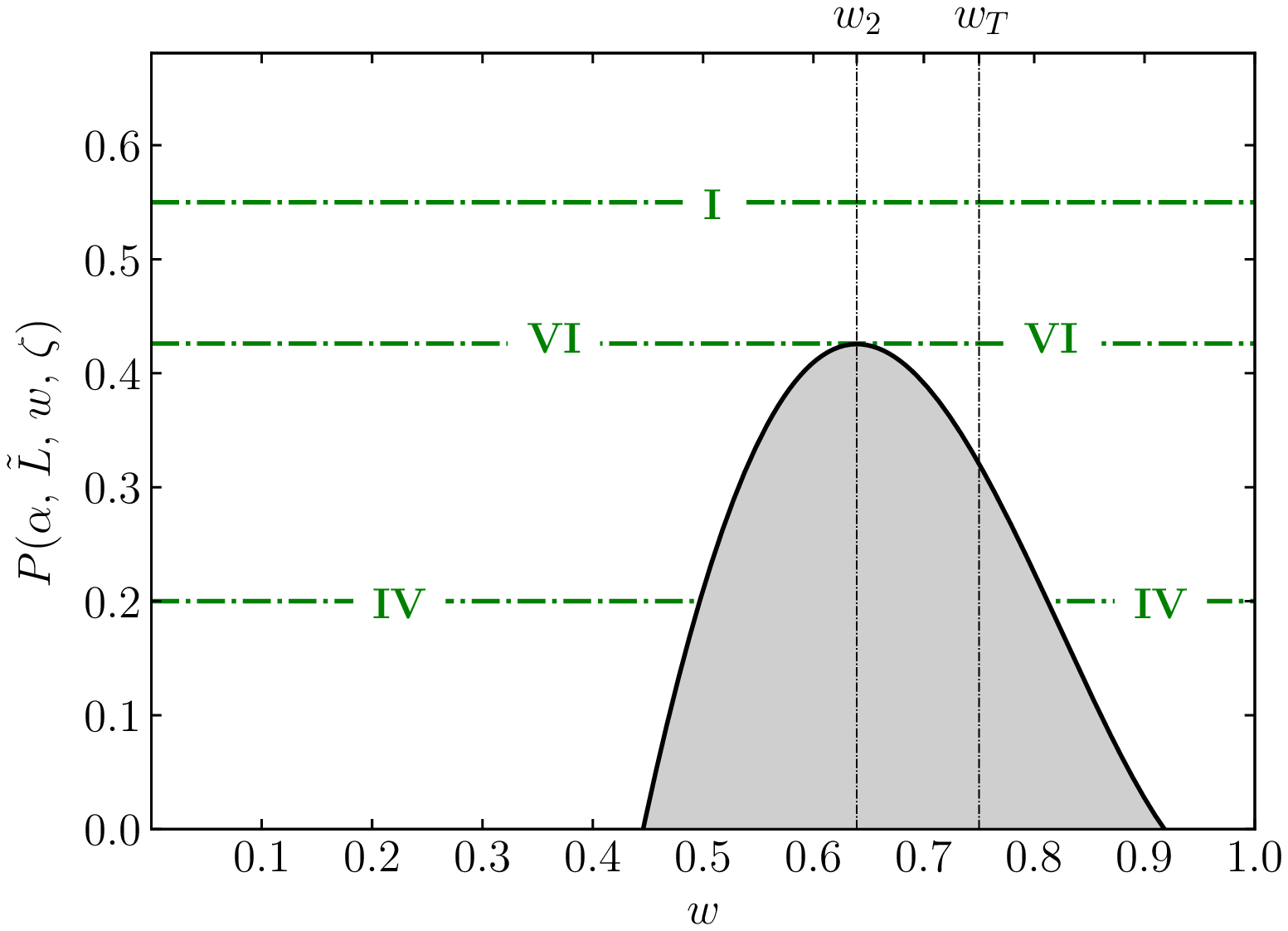}\label{Sol:drphi:Class:Fig:otherPoly:zeta1}}
\caption{The polynomial for (a) lightlike and (b) spacelike geodesics versus the radius $w$. Grey areas represent physically forbidden zones. For the spacelike geodesics a physically forbidden zone occurs only when $\tilde{L} > \tilde{L}_\text{zero}$. Note that for spacelike geodesics only real energies are admitted.}
\label{Sol:drdphi:Class:Fig:otherPoly}
\end{figure}

The orbits for the spacelike geodesics are shown in Fig. \ref{Sol:drphi:Class:Fig:otherPoly:zeta1}. Here we note that although Fig. \ref{Sol:drdphi:Class:Fig:lcrit:z1} shows three characteristic angular momenta, only one of them is relevant when considering only real values for the energies, i.e. $\bar{\epsilon}^2 > 0$. For $\tilde{L} < \tilde{L}_\text{zero}$ only transit orbits are possible because the polynomial $P$ is negative in the full coordinate range $w \in [0,1]$. Only for $\tilde{L} > \tilde{L}_\text{zero}$ a physically forbidden zone arises around $w_2$, which leads in addition to escape orbits.

With the above definitions of the polynomial and its extremal points, as well as the discussion on the characteristic angular momenta for spacelike and timelike geodesics, the orbit types can now be defined. We will apply this definition in the parameter plots in \autoref{Sol:drdphi:Class:Fig:ParamPlot}. In Table (\ref{Sol:drdphi:Class:Tab:Types} a summary of the orbit types is presented.

\begin{itemize}
\item Type \textbf{\RN{1}} (TO):
\begin{equation}
\bar{\epsilon}^2 > P(\bar{\alpha}_+, \tilde{L}, w, \zeta),~~ \forall w \in [0,1].
\end{equation}

Type \textbf{\RN{1}} orbits consist only of transit orbits (TO). These arise for timelike, lightlike and spacelike geodesics.

\item Type \textbf{\RN{2}} (EO, TWEO):
\begin{equation}
P_1(\bar{\alpha}_+,\; \zeta) < \bar{\epsilon}^2 < \min(P_{\text{max}}(\bar{\alpha}_+, \tilde{L}, \zeta), P_0(\bar{\alpha}_+,\; \zeta)).\,
\end{equation}

Type \textbf{\RN{2}} orbits contain both escape orbits (EO) and two world escape orbits (TWEO). These orbits never reach the asymptotic region $w=0$. They arise only for timelike geodesics.

\item Type \textbf{\RN{3}} (BO, EO, TWEO):
\begin{equation}
\tilde{L}_\text{crit} < \tilde{L} \leqslant \tilde{L}_\text{swap} :\; P_{\text{min}}(\bar{\alpha}_+, \tilde{L}, \zeta) \leqslant \bar{\epsilon}^2 \leqslant P_{\text{max}}(\bar{\alpha}_+, \tilde{L}, \zeta).
\end{equation}

Type \textbf{\RN{3}} orbits contain also bound orbits (BO) besides the escape orbits (EO) and two world escape orbits (TWEO). These orbits never reach the asymptotic region $w=0$. They arise only for timelike geodesics.
 
\item Type \textbf{\RN{4}} (EO, TWEO):
\begin{equation}
\tilde{L} > \tilde{L}_\text{swap}:\; P_0(\bar{\alpha}_+,\; \zeta) < \bar{\epsilon}^2 < P_{\text{max}}(\bar{\alpha}_+, \tilde{L}, \zeta)
\end{equation}

Type \textbf{\RN{4}} orbits contain both escape orbits (EO) and two world escape orbits (TWEO).In contrast to Type \textbf{\RN{2}} orbits there are also escape orbits (EO), which reach the asymptotic region $w=0$. They arise for timelike, lightlike and spacelike geodesics.

\item Type \textbf{\RN{5}} (UCO, BO, EO):
\begin{equation}
\tilde{L} > \tilde{L}_\text{crit}:\; \bar{\epsilon}^2 = P_{\text{max}}(\bar{\alpha}_+, \tilde{L}, \zeta)
\end{equation}

In Type \textbf{\RN{5}} orbits the particle energy corresponds to the maximum of the polynomial $P$ for timelike geodesics. Here in addition to the orbits of Type \textbf{\RN{3}} an unstable circular orbit (UCO) is present.

\item Type \textbf{\RN{6}} (UCO, TWEO, EO):
\begin{equation}
\bar{\epsilon}^2 = P_{\text{max}}(\bar{\alpha}_+, \tilde{L}, \zeta)
\end{equation}

For timelike geodesics with $\tilde{L} > \tilde{L}_\text{crit}$ and for spacelike geodesics with $\tilde{L} > \tilde{L}_\text{zero}$ there is in addition to the orbits present in Type \textbf{\RN{4}} 
an unstable circular orbit (UCO). For lightlike geodesics this type is always present.

\item Type \textbf{\RN{7}} (UCO, EO):
\begin{equation}
\tilde{L} = \tilde{L}_\text{crit}:\; \epsilon^2 = P_{\text{max}}(\bar{\alpha}_+, \tilde{L}, \zeta) = P_{\text{min}}(\bar{\alpha}_+, \tilde{L}, \zeta)
\end{equation}

Type \textbf{\RN{7}} is found for $\tilde{L} = \tilde{L}_\text{crit}$. In this case the zeros of $P'$ coincide, $w_2 = w_3$, thus $P_{\text{max}}(\bar{\alpha}_+, \tilde{L}, \zeta) = P_{\text{min}}(\bar{\alpha}_+, \tilde{L}, \zeta)$. When the particle energy corresponds to the potential at $w_2 = w_3$, an unstable circular orbit (UCO) arises in addition to the orbits of Type \textbf{\RN{2}}.
\end{itemize}

\begin{table}[H]
\begin{center}
\begin{tabular}{|lcccl|}\hline
Type & $\zeta$ & Zeros  & Range of $x$ & Orbit \\
\hline\hline
\RN{1} & $-1$, 0, 1 & 0 &
\begin{pspicture}(-3,-0.2)(3,0.2)
\psline[linewidth=0.5pt]{|->}(-3,0)(3,0)
\psline[linewidth=0.5pt,doubleline=true](0,-0.2)(0,0.2)
\psline[linewidth=1.2pt]{-}(-3,0)(2.9,0)
\end{pspicture}
& TO \\ 

\hline
\RN{2}$_a$ & $-1$ & 1 &
\begin{pspicture}(-3,-0.2)(3,0.2)
\psline[linewidth=0.5pt]{|->}(-3,0)(3,0)
\psline[linewidth=0.5pt,doubleline=true](0,-0.2)(0,0.2)
\psline[linewidth=1.2pt]{-*}(-3,0)(-0.5,0)
\end{pspicture}
& EO \\

\hline
\RN{2}$_b$ & $-1$ & 1 &
\begin{pspicture}(-3,-0.2)(3,0.2)
\psline[linewidth=0.5pt]{|->}(-3,0)(3,0)
\psline[linewidth=0.5pt,doubleline=true](0,-0.2)(0,0.2)
\psline[linewidth=1.2pt]{-*}(-3,0)(0.5,0)
\end{pspicture}
& TWEO \\

\hline
\RN{3}$_a$ & $-1$ & 3 &
\begin{pspicture}(-3,-0.2)(3,0.2)
\psline[linewidth=0.5pt]{|->}(-3,0)(3,0)
\psline[linewidth=0.5pt,doubleline=true](0,-0.2)(0,0.2)
\psline[linewidth=1.2pt]{*-*}(1.0,0)(2.5,0)
\psline[linewidth=1.2pt]{-*}(-3,0)(-0.5,0)
\end{pspicture}
  & EO, BO \\
  
\hline
\RN{3}$_b$ & $-1$ & 3 &
\begin{pspicture}(-3,-0.2)(3,0.2)
\psline[linewidth=0.5pt]{|->}(-3,0)(3,0)
\psline[linewidth=0.5pt,doubleline=true](0,-0.2)(0,0.2)
\psline[linewidth=1.2pt]{*-*}(1,0)(2.5,0)
\psline[linewidth=1.2pt]{-*}(-3,0)(0.5,0)
\end{pspicture}
  & TWEO, BO \\
  
\hline
\RN{4}$_a$ & $-1$, 0, 1 & 2 &
\begin{pspicture}(-3,-0.2)(3,0.2)
\psline[linewidth=0.5pt]{|->}(-3,0)(3,0)
\psline[linewidth=0.5pt,doubleline=true](0,-0.2)(0,0.2)
\psline[linewidth=1.2pt]{*-}(1.5,0)(2.9,0)
\psline[linewidth=1.2pt]{-*}(-3,0)(-0.5,0)
\end{pspicture}
  & EO \\
  
\hline
\RN{4}$_b$ & $-1$, 0, 1 & 2 &
\begin{pspicture}(-3,-0.2)(3,0.2)
\psline[linewidth=0.5pt]{|->}(-3,0)(3,0)
\psline[linewidth=0.5pt,doubleline=true](0,-0.2)(0,0.2)
\psline[linewidth=1.2pt]{*-}(1.5,0)(2.9,0)
\psline[linewidth=1.2pt]{-*}(-3,0)(0.5,0)
\end{pspicture}
  & TWEO, EO \\
  
\hline
\RN{5} & $-1$ & 3 &
\begin{pspicture}(-3,-0.2)(3,0.2)
\psline[linewidth=0.5pt]{|->}(-3,0)(3,0)
\psline[linewidth=0.5pt,doubleline=true](0,-0.2)(0,0.2)
\psline[linewidth=1.2pt]{o-*}(0.5,0)(2.5,0)
\psline[linewidth=1.2pt]{-o}(-3,0)(0.5,0)
\end{pspicture}
  & TWEO, UCO, BO \\

\hline
\RN{6} & $-1$, 0, 1 & 2 &
\begin{pspicture}(-3,-0.2)(3,0.2)
\psline[linewidth=0.5pt]{|->}(-3,0)(3,0)
\psline[linewidth=0.5pt,doubleline=true](0,-0.2)(0,0.2)
\psline[linewidth=1.2pt]{o-}(0.5,0)(2.9,0)
\psline[linewidth=1.2pt]{-o}(-3,0)(0.5,0)
\end{pspicture}
  & TWEO, UCO, EO \\
  
  \hline
\RN{7} & $-1$ & 1 &
\begin{pspicture}(-3,-0.2)(3,0.2)
\psline[linewidth=0.5pt]{|->}(-3,0)(3,0)
\psline[linewidth=0.5pt,doubleline=true](0,-0.2)(0,0.2)
\psline[linewidth=1.2pt]{-o}(-3,0)(0.5,0)
\end{pspicture}
& TWEO, UCO \\
\hline\hline
\end{tabular}

\caption{Orbit types for timelike ($\zeta = -1$), lightlike ($\zeta = 0$) and spacelike ($\zeta = 1$) geodesics. Circles represent zeros of $(d\bar{r} / d\phi)^2$, i.e., turning points; thick lines represent possible motion in the range of the radial coordinate $x$, which ranges from $\frac{1}{2}$ to $\infty$. The wormhole throat $x_T$ is indicated by the vertical double line at the center. Open circles indicate potential maxima or saddle points, where UCOs reside.}
\label{Sol:drdphi:Class:Tab:Types}
\end{center}
\end{table}

We consider the possibility of two world orbits of timelike geodesics, i.e., orbits of particles which cross the wormhole throat, particularly interesting. Transit orbits (TO) represent trivial two world orbits. To find other types of two world orbits we have to study the extrema $w_1$, $w_2$ and $w_3$ with respect to the throat location $w_T \in (0.5,1)$ with $\zeta = -1$ in the following section.

Since $\lim_{\tilde{L} \rightarrow \infty} w_3(\tilde{L},\, \bar{\alpha}_+,\, \zeta) = 0$ and $\lim_{\tilde{L} \rightarrow \infty} w_2(\tilde{L},\, \bar{\alpha}_+,\, \zeta) = \frac{1}{2}$ we note that $w_3 < w_2 < \frac{1}{2}$ for any parameters. Therefore bound orbits occur only in one world.They cannot cross the throat. However, escape orbits can cross the throat. Their energy only has to exceed the value of the potential at the throat $P(\bar{\alpha}_{+}, \tilde{L}, w = w_T, \zeta) := P_T(\bar{\alpha}_{+}, \tilde{L}, \zeta)$, which we refer to as wormhole barrier. In order to differentiate between two world and single world orbits we put a subscript to the orbit types, where both are possible. Single world orbits never exceed the wormhole barrier $\bar{\epsilon}^2 \leqslant P_T(\bar{\alpha}_{+}, \tilde{L}, \zeta)$. They carry the subscript $a$. Two world orbits can cross the wormhole barrier, since $\bar{\epsilon}^2 > P_T(\bar{\alpha}_{+}, \tilde{L}, \zeta)$. They carry the subscript $b$. \\

Clearly, the extremal values for lightlike geodesics $w_1, w_2$ and $w_3$ are independent of $\bar{\alpha}_+$ and $\tilde{L}$. The extremal values for spacelike geodesics depend on both parameters again, but lead to a behavior different from timelike geodesics. Here it can happen for fixed $\bar{\alpha}_+$ and increasing $\tilde{L}$, that the inequality $w_T < w_2$ changes to $w_2 < w_T$ (i.e., the same inequality as for timelike geodesics). A closer examination of this circumstance shows that this happens, however, only when $\tilde{L} = \tilde{L}_\text{zero}$. Due to the restriction to $\bar{\epsilon}^2 > 0$, we only encounter the inequality $w_2 < w_T$ in our discussion. Therefore the geodesics of type \RN{4}$_b$ and \RN{6} orbits, which temporarily do or do not cross the throat, cannot originate in the other asymptotic region, respectively.

By plotting all extremal values of the polynomial together with the end points $P_0$ and $P_1$, a parametric plot $\bar{\epsilon}^2$ over $\tilde{L}$ can be created, as shown in \autoref{Sol:drdphi:Class:Fig:ParamPlot} for some parameter set. The parametric plot exhibits the dependence of the possible orbit types on the energy and the angular momentum. In the parameter plots for fixed $\zeta$ a variation of $\bar{\alpha}_+$ will only change the size of the regions but not their overall structure. Therefore it is sufficient to consider only one set of parameters to assign the orbits types.

For timelike geodesics the lowest values of the characteristic angular momenta $\tilde{L}_\text{crit}$ 
and $\tilde{L}_\text{swap}$ are clearly seen in \autoref{Sol:drdphi:Class:Fig:ParamPlot:z-1}. Furthermore \autoref{Sol:drdphi:Class:Fig:ParamPlot:z-1} and\autoref{Sol:drdphi:Class:Fig:ParamPlot:z1} show the wormhole barrier $P_T$ as a blue line (dotted), which divides the regions for the orbit types into two world and single world orbits. The location of the wormhole barrier curve depends on the wormhole parameter $\bar{\alpha}_+$. Increasing the value of $\bar{\alpha}_+$ leads to a smaller inclination, and thus two world orbits can occur for smaller energies.

As demonstrated in \autoref{Sol:drdphi:Class:Fig:ParamPlot:z0}, only a single parametric plot exists for lightlike geodesics, since there is no dependence on the wormhole parameter $\bar{\alpha}_+$. In contrast the wormhole barrier does depend on $\bar{\alpha}_+$, as demonstrated in the figure as well. Here a larger wormhole parameter $\bar{\alpha}_+$ gives rise to two world orbits for smaller energies $\bar{\epsilon}^2$. 

Restricting the analysis to $\bar{\epsilon}^2 > 0$, we show the parametric plot of spacelike geodesics in \autoref{Sol:drdphi:Class:Fig:ParamPlot:z1}, which turns out rather simple as well. If we were to consider the parametric plots for negative energy square $\bar{\epsilon}^2 < 0$ as well, we would find structures similar to \autoref{Sol:drdphi:Class:Fig:ParamPlot:z-1}. This would indicate the existence of bound orbits for particles with imaginary energy. 

\subsection{Solution for the \boldmath{$(d\bar{r} / dt)$}-motion}
\label{Sol:drdt:Sec}
In order to solve for the $(d\bar{r}/dt)$-motion, \autoref{Sol:Eq:drdt2} is transformed as follows

\begin{equation}
\left(\frac{d\bar{r}}{dt}\right)^2 = \frac{L^2}{\epsilon^2 \left(1 + \frac{\eta}{2 \bar{r}}\right)^8 \bar{r}^4} \left(\frac{d\bar{r}}{d\phi}\right)^2\,.
\end{equation}

We then apply again the coordinate transformation $\bar{r} = \eta (x - \frac{1}{2}),\; d\bar{r} = \eta dx$, and introduce again $\tilde{L}$ and $\bar{\epsilon}$ together with the new time coordinate $hat{t} = \frac{t}{\eta}$. This yields

\begin{equation}
\label{Sol:drdt:Eq:final_dform}
\left(\frac{dx}{d\hat{t}}\right)^2 = \frac{\tilde{L}^2}{16 \bar{\epsilon}^2}\frac{(2 x - 1)^4}{x^8} \left(\frac{dx}{d\phi}\right)^2\,.
\end{equation}

\noindent Also we set $(dx/d\phi)^2 = P_4(x)$. Integrating \autoref{Sol:drdt:Eq:final_dform} we get

\begin{equation}
\hat{t} - \hat{t}_0 = \frac{4 \bar{\epsilon}}{\tilde{L}} \int\limits_{x_0}^x \frac{x^4}{(2 x - 1)^2} \frac{dx}{\sqrt{P_4(x)}}\,.
\end{equation}

For $P_4(x)$ we already know the solution in terms of the Weierstra\ss \ function. When using the Weierstra\ss \ function we can later apply the addition theorems of this function in order to solve the $(d\bar{r}/dt)$-equation. Therefore we employ $x = \frac{b_3}{4 y - \frac{b_2}{3}} + x_1$ to transform $P_4(x) \rightarrow P_3(y)$. With partial fractions this new integral can be written as

\begin{equation}
\hat{t} - \hat{t}_0 = \frac{4 \bar{\epsilon}}{\tilde{L}} \int\limits_{y_0}^y \left[k_0 + \sum\limits_{i=1}^2 \frac{a_i}{y - p_i} + \sum\limits_{i=1}^2 \frac{b_i}{(y - p_i)^2}\right] \frac{dy}{\sqrt{P_3(y)}}\,,
\end{equation}

\noindent with poles

\begin{eqnarray}
p_1 &=& \frac{b_2}{12}\,, \notag \\
p_2 &=& \frac{1}{12} \frac{2 b_2 x_1 - b_2 - 6 b_3}{2 x_1 - 1}\,,
\end{eqnarray}

\noindent and coefficients

\begin{eqnarray}
k_0 &=& \frac{x_1^4}{4 x_1^2 - 4 x_1 + 1}\,, \notag \\
a_1 &=& \frac{b_3^2}{64}\,, \notag \\
a_2 &=&  -\frac{b_3}{16}\frac{4 x_1 - 1}{(2 x_1 - 1)^3}\,, \\
b_1 &=& \frac{b_3}{16} (2 x_1 + 1)\,, \notag \\
b_2 &=&  \frac{1}{64} \frac{b_3^2}{(2 x_1 - 1)^4}\,. \notag
\end{eqnarray}

The integration of the time coordinate $\hat{t}$ can be reduced to two different problems to be solved. One problem is the integration of an elliptic integral of the third kind, for further discussion defined as $I_1$, and the other one is the integration of an elliptic integral with a double pole $I_2$

\begin{equation}
I_1 = \int\limits_{y_0}^y \frac{1}{y - p} \frac{dy}{\sqrt{P_3(y)}},~~ I_2 = \int\limits_{y_0}^y \frac{1}{(y - p)^2} \frac{dy}{\sqrt{P_3(y)}}\,.
\end{equation}

Those two integrals can indeed be solved by using addition theorems of the Weierstra\ss \ functions, thus yielding the solution for the time coordinate $\hat{t}$

\begin{equation}
\hat{t}(v)  = \frac{4 \bar{\epsilon}}{\tilde{L}} \left(k_0 (v - v_0) + a_1 I_1 + a_2 I_2 + b_1 I_1 + b_2 I_2\right) + \hat{t}_0\,,
\end{equation}

\noindent where ${v_p}_i$ is a Weierstra\ss \ transformed pole $p_i$ with the relation $p_i = \wp({v_p}_i)$. The solution of $I_1$ and $I_2$ will be presented in the next two paragraphs.

\paragraph{Solution of the elliptic integrals of third kind $I_1$} \mbox{} \\
Let us begin by introducing yet another coordinate transformation which will simplify the integration. With $y = \wp(v)$, which solves the Weierstra\ss \ differential equation (\ref{Sol:drdphi:Eq:dydphi}) and $dy = dv \sqrt{P_3(y)}$, $I_1$ becomes

\begin{equation}
\label{Sol:drdt:I1:Eq:I1sub}
I_1 = \int\limits_{v_0}^v \frac{dv}{\wp(v) - p} = \int\limits_{v_0}^v \frac{dv}{\wp(v) - \wp(v_p)}\,.
\end{equation}

\noindent where in the second term $p = \wp(v_p)$ redefines the pole $p$ using the Weierstra\ss \ function with a corresponding input parameter $v_p$. With the constant $\wp'(v_p)$, $I_1$ can be written as

\begin{equation}
I_1 = \frac{1}{\wp'(v_p)} \int\limits_{v_0}^v \frac{\wp'(v_p)}{\wp(v) - \wp(v_p)} dv\,.
\end{equation}

In this form we can apply an addition theorem for the Weierstra\ss \ functions in the integrand \cite{Lawden:1989}

\begin{equation}
\label{Sol:drdt:I1:Eq:addtheo}
\frac{\wp'(y)}{\wp(x) - \wp(y)} = -\zeta(x + y) + \zeta(x - y) + 2 \zeta(y)\,,
\end{equation}

\noindent where $\zeta(z) = -\int \wp(z) dz$ is the Weierstra\ss \ $\zeta$-function. Inserting \autoref{Sol:drdt:I1:Eq:addtheo} into $I_1$ gives

\begin{equation}
I_1 = \frac{1}{\wp'(v_p)} \int\limits_{v_0}^v -\zeta(v + v_p) + \zeta(v - v_p) + 2 \zeta(v_p) dv\,,
\end{equation}

\noindent which can be trivially integrated with the relation $\frac{d}{dz} \ln \sigma(z) = \zeta(z)$, where $\sigma(z)$ is the Weierstra\ss \ $\sigma$-function.

\begin{equation}
I_1 = \frac{1}{\wp'(v_p)} \left[2 \zeta(v_p)(v - v_p) + \ln \frac{\sigma(v - v_p)}{\sigma(v + v_p)} - \ln \frac{\sigma(v_0 - v_p)}{\sigma(v_0 + v_p)}\right]
\end{equation}

\paragraph{Solution of the elliptic integrals of type $I_2$} \mbox{} \\
The elliptic integrals of type $I_2$ need some more discussion as compared to $I_1$. $I_2$ is rewritten in the same way as $I_1$, see \autoref{Sol:drdt:I1:Eq:I1sub},

\begin{equation}
\label{Sol:drdt:Eq:0}
I_2 = \int\limits_{v_0}^v \frac{dv}{(\wp(v) - p)^2} = \int\limits_{v_0}^v \frac{dv}{(\wp(v) - \wp(v_p))^2}\,.
\end{equation}

\noindent Taking \autoref{Sol:drdt:I1:Eq:addtheo}

\begin{equation}
\label{Sol:drdt:Eq:1}
\zeta(u + v) + \zeta(u - v) - 2 \zeta(u) = \frac{\wp'(u)}{\wp(u) - \wp(v)}
\end{equation}

\noindent and swapping $u$ and $v$ and using the relation $\zeta(-z) = - \zeta(z)$ we get

\begin{equation}
\label{Sol:drdt:Eq:2}
\zeta(u + v) - \zeta(u - v) - 2 \zeta(v) = \frac{\wp'(v)}{\wp(u) - \wp(v)}\,.
\end{equation}

\noindent Adding \autoref{Sol:drdt:Eq:1} and \autoref{Sol:drdt:Eq:2} yields

\begin{equation}
\label{Sol:drdt:Eq:3}
\zeta(u + v) = \zeta(u) + \zeta(v) + \frac{1}{2} \frac{\wp'(u) - \wp'(v)}{\wp(u) - \wp(v)}\,.
\end{equation}

\noindent Differentiating \autoref{Sol:drdt:Eq:3} leads to \autoref{Sol:drdt:Eq:4}. Here $u$ and $v$ can be swapped again, resulting in \autoref{Sol:drdt:Eq:5}.

\begin{equation}
\label{Sol:drdt:Eq:4}
\wp(u + v) = \wp(u) - \frac{1}{2} \frac{\wp''(u)}{\wp(u) - \wp(v)} + \frac{1}{2} \frac{(\wp'(u) - \wp'(v)) \wp'(u)}{(\wp(u) - \wp(v))^2}
\end{equation}

\begin{equation}
\label{Sol:drdt:Eq:5}
\wp(u + v) = \wp(v) + \frac{1}{2} \frac{\wp''(v)}{\wp(u) - \wp(v)} - \frac{1}{2} \frac{(\wp'(u) - \wp'(v)) \wp'(v)}{(\wp(u) - \wp(v))^2}
\end{equation}

\noindent With \autoref{Sol:drdt:Eq:5} the integrand of \autoref{Sol:drdt:Eq:0} can be rewritten as

\begin{equation}
\label{Sol:drdt:Eq:6}
\frac{1}{(\wp(v) - \wp(v_p))^2} = \frac{1}{\wp'(v_p)^2} \left[2 \wp(v + v_p) - 2 \wp(v_p) - \frac{\wp''(v_p)}{\wp(v) - \wp(v_p)}\right] + \frac{1}{\wp(v_p)} \frac{\wp'(v)}{(\wp(v) - \wp(v_p))^2}\,.
\end{equation}

Knowing that $\int \wp(v) dv = -\zeta(v)$ and that

\begin{equation}
\label{Sol:drdt:Eq:7}
\int \frac{\wp'(z)}{(\wp(z) - \wp(v_p))^2} dz = \int \frac{du}{u^2} = -\frac{1}{u} = -\frac{1}{\wp(z) - \wp(v_p)}
\end{equation}

\noindent can be integrated by substituting $u = \wp(v) - \wp(v_p)$ and $dv = \frac{du}{\wp'(v)}$, $I_2$ can be integrated,

\begin{equation}
\label{Sol:drdt:Eq:prefinal}
I_2 = - \frac{\wp''(v_p)}{\wp'(v_p)^2} I_1 - \frac{1}{\wp'(v_p)^2} \left[2 \zeta(v + v_p) + 2 \zeta(v_0 + v_p) + 2 \wp(v_p) (v - v_0) + \frac{\wp'(v_p)}{\wp(v) - \wp(v_p)} + \frac{\wp'(v_p)}{\wp(v_0) - \wp(v_p)} \right]\,.
\end{equation}

\noindent Finally, applying \autoref{Sol:drdt:Eq:3} reduces the previously calculated solution of $I_2$ to

\begin{equation}
I_2 = - \frac{\wp''(v_p)}{\wp'(v_p)^2} I_1 - \frac{1}{\wp'(v_p)^2} \left[2 \wp(v_p) (v - v_0) + 2 (\zeta(v) + \zeta(v_0)) + \frac{\wp'(v)}{\wp(v) - \wp(v_p)} + \frac{\wp'(v_0)}{\wp(v_0) - \wp(v_p)} \right]\,.
\end{equation}

\section{The orbits}
\label{Orbits:Sec}
With the solution for $(d\bar{r}/d\phi)$ we found the analytical solution for the motion of particles and light in the equatorial plane $\vartheta = \frac{\pi}{2}$. Before plotting some of the orbits in the wormholes spacetimes, we need to evaluate the isometric embedding of the traversable wormhole solutions. To this end we transform the metric $ds^2$ from \autoref{JNWW:Eq:Metrik} with $dt = 0$ into cylindrical coordinates

\begin{eqnarray}
ds^2 &\overset{!}{=}& dX^2 + dY^2 + dz^2 \notag \\
&=& d\rho^2 + dz^2 + \rho^2 d\phi^2 \notag \\
&=& \left[\left(\frac{d\rho}{d\bar{r}}\right)^2 + \left(\frac{dz}{d\bar{r}}\right)^2\right] d\bar{r}^2 + \rho^2 d\phi^2.
\label{euclid}
\end{eqnarray}

\noindent The coordinate $\rho$ can be directly read from the original metric and has the form

\begin{eqnarray}
\label{Orbits:Eq:pre_rho}
\rho(\bar{r}) = \alpha_+ \left(\bar{r} - \frac{\eta^2}{4 \bar{r}}\right) + \alpha_- \left(\bar{r} + \eta + \frac{\eta^2}{4 \bar{r}}\right)\,.
\end{eqnarray}

Transforming to the radial coordinate $x = \frac{\bar{r}}{\eta} + \frac{1}{2}$, 

\begin{equation}
\tilde{\rho}(x) = \frac{2 (\bar{\alpha}_+ + 1) x^2 - 2 \bar{\alpha}_+ x}{2 x - 1}\,,
\end{equation}

\noindent where $\tilde{\rho}(x) = \frac{\rho(\bar r)}{\eta \alpha_-}$, we find for the wormhole throat

\begin{equation}
\label{Orbits:Eq:ThroatRadius}
\tilde{\rho}(x_T) = \tilde{\rho}_T(\bar{\alpha}_+) = 1 + \sqrt{1 - \bar{\alpha}_+^2}\,.
\end{equation}

\noindent This shows that the circumferential throat radius satisfies $\tilde{\rho}_T > 1$ for all values of the wormhole parameter $\bar{\alpha}_+$.

To obtain the coordinate $z$ we equate the coefficient of $d\bar{r}^2$ in \autoref{euclid} with the one of the metric in \autoref{JNWW:Eq:Metrik},

\begin{equation}
\label{Orbits:Eq:pre_zbar}
\left[\left(\frac{d\rho}{d\bar{r}}\right)^2 + \left(\frac{dz}{d\bar{r}}\right)^2\right] = \left[\alpha_+ \frac{\left(1-\frac{\eta}{2 \bar{r}}\right)}{\left(1+\frac{\eta}{2 \bar{r}}\right)} + \alpha_-\right]^2\left(1+\frac{\eta}{2 \bar{r}}\right)^4.
\end{equation}

Transforming again to the radial coordinate $x = \frac{\bar{r}}{\eta} + \frac{1}{2}$ we obtain

\begin{equation}
\label{Orbits:Eq:zbar}
\tilde{z}(x) - \underbrace{\tilde{z}(x_T)}_{:=0} = \mathop{\mathlarger{\mathlarger{\mathlarger{\int}}}}\limits_{x_T}^{x}\frac{\pm\sqrt{16\left[\bar{\alpha}_+ \left(1 - \dfrac{1}{x}\right) +  1\right]^2 x^4 - \Big[(2 x - 1)^2 (\bar{\alpha}_+ + 1) + \bar{\alpha}_+ - 1\Big]^2}}{(2 x - 1)^2}\;dx\,
\end{equation}

\noindent with $\tilde{z}(x) = \frac{z(\bar r)}{\eta \alpha_-}$. This integral is evaluated numerically. \\

\begin{figure}[H]
\centering
\subfloat[$\bar{\alpha}_+ = 0.100,\, \tilde{z}_\text{lim} = 60$]{\includegraphics[width=0.26\linewidth]{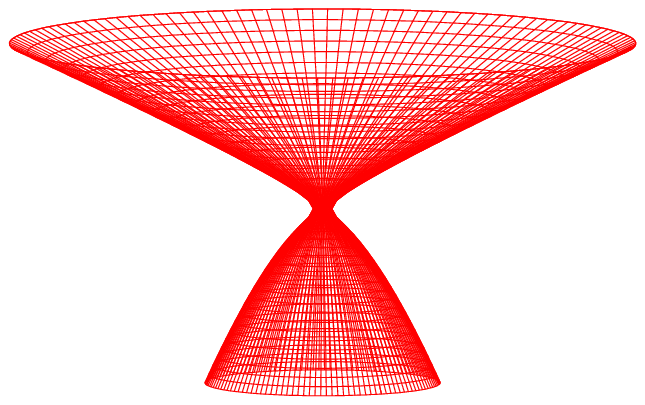}}
\subfloat[$\bar{\alpha}_+ = 0.800,\, \tilde{z}_\text{lim} = 60$]{\includegraphics[width=0.25\linewidth]{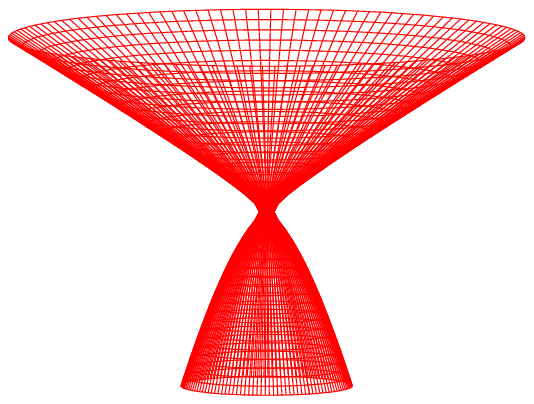}}
\subfloat[$\bar{\alpha}_+ = 0.970,\, \tilde{z}_\text{lim} = 60$]{\includegraphics[width=0.25\linewidth]{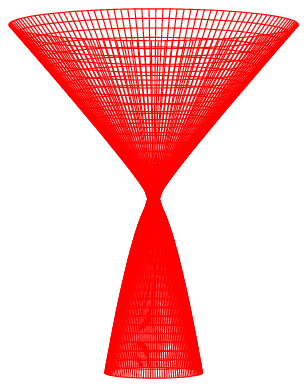}}

\subfloat[$\bar{\alpha}_+ = 0.100,\, \tilde{z}_\text{lim} = 7$]{\includegraphics[width=0.26\linewidth]{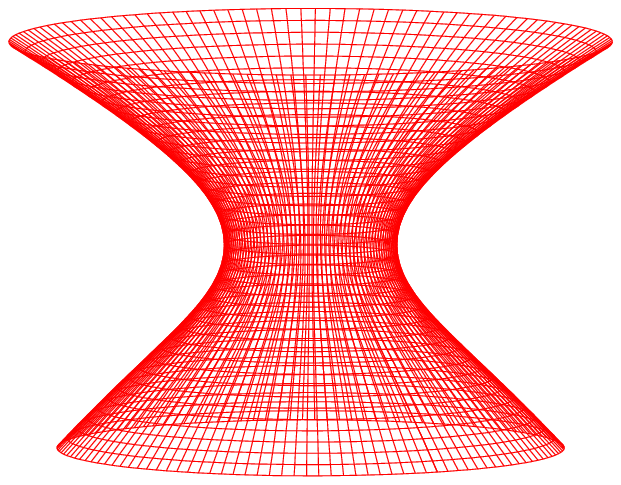}}
\subfloat[$\bar{\alpha}_+ = 0.800,\, \tilde{z}_\text{lim} = 7$]{\includegraphics[width=0.25\linewidth]{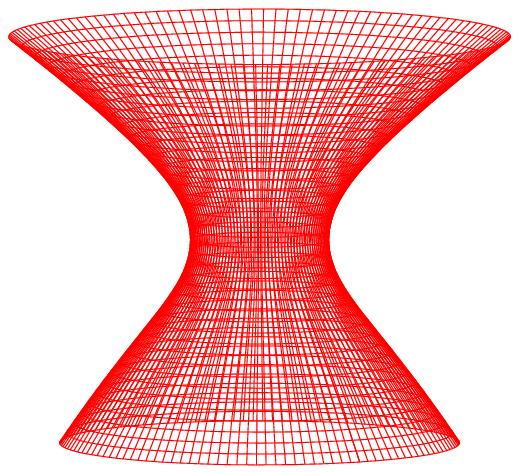}}
\subfloat[$\bar{\alpha}_+ = 0.970,\, \tilde{z}_\text{lim} = 7$]{\includegraphics[width=0.25\linewidth]{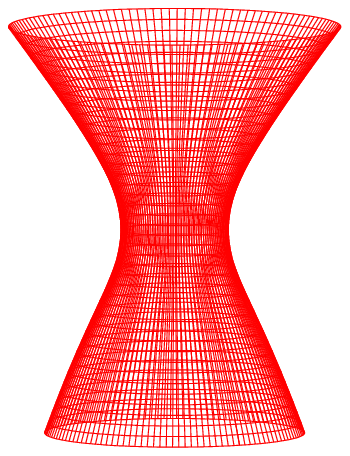}}

\caption{Isometric embeddings of the wormhole metric for different wormhole parameters $\bar{\alpha}_+$. In all figures the shown intervals are the same for all axes and defined by the value of $\tilde{z}_\text{lim}$, e.g., $\tilde{z} \in \left[-\tilde{z}_\text{lim},\, \tilde{z}_\text{lim}\right]$. Thus the choice of $\tilde{z}_\text{lim}$ determines the detail of the figures. Note, that figures (d)-(f) represent zooms of the throat region of figures (a)-(c).}
\label{Orbits:Fig:WormholeManifold}
\end{figure}

With $\tilde{z}(x)$ and $\tilde{\rho}(x)$ at hand, we now illustrate the shape of the wormholes 
in \autoref{Orbits:Fig:WormholeManifold} for different choices of the parameter $\bar{\alpha}_+$.

\begin{figure}[H]
\centering
\subfloat[\RN{3}$_a$-orbit (BO), $\bar{\epsilon}^2 = 1.66,\, \tilde{L} = 3.00,
\, \bar{\alpha}_+ = 0.50,\, \zeta = -1$]{\label{Orbits:Fig:Orbits1:1}\includegraphics[width=0.42\linewidth]{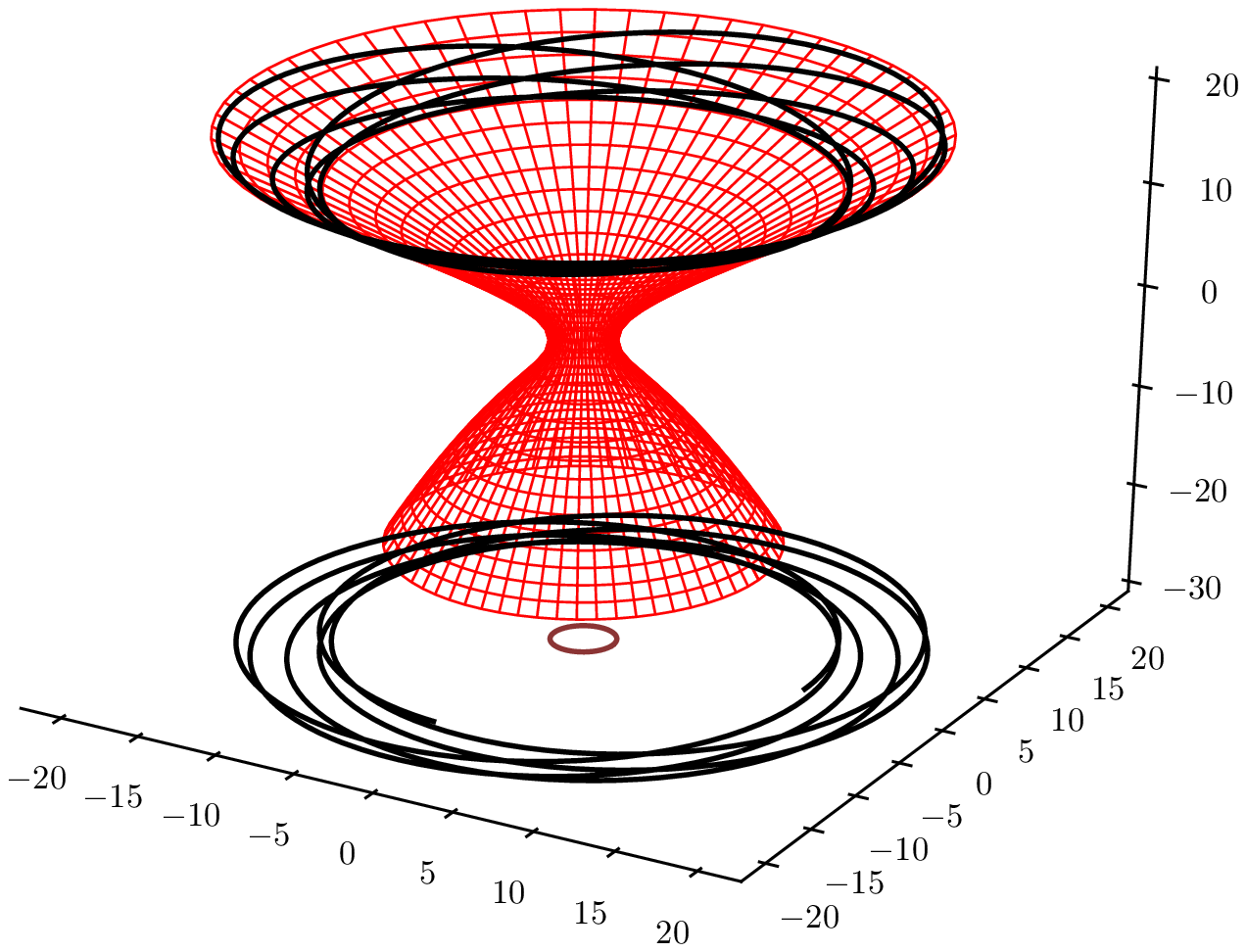}}
\subfloat[\RN{2}$_b$-orbit (TWEO), $\bar{\epsilon}^2 = 2.20,\, \tilde{L} = 2.00,\, \bar{\alpha}_+ = 0.50,\, \zeta = -1$]{\label{Orbits:Fig:Orbits1:2}\includegraphics[width=0.42\linewidth]{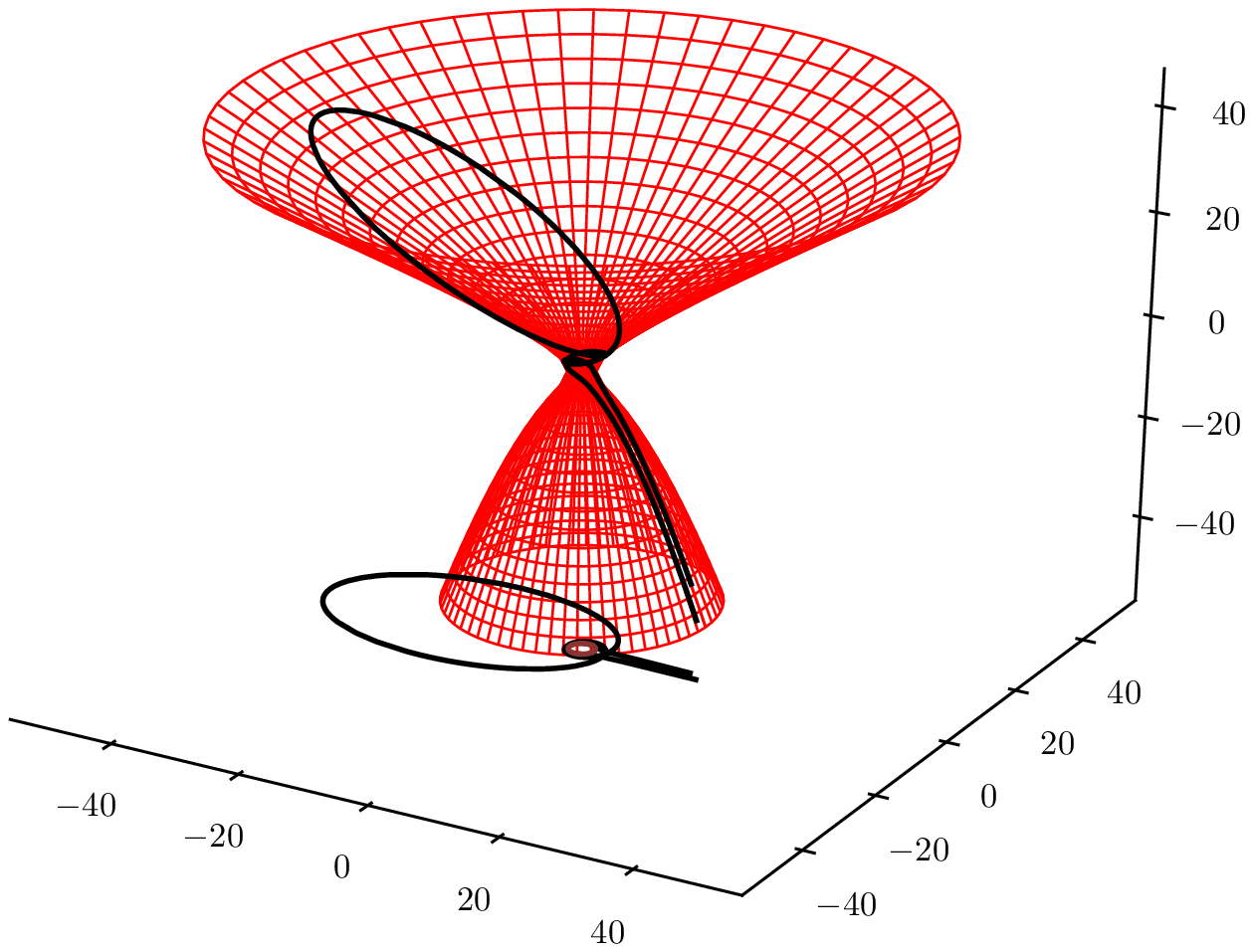}} \\
\subfloat[\RN{2}$_b$-orbit (TWEO), $\bar{\epsilon}^2 = 3.25,\, \tilde{L} = 2.80,\, \bar{\alpha}_+ = 0.90,\, \zeta = -1$]{\label{Orbits:Fig:Orbits1:3}\includegraphics[width=0.42\linewidth]{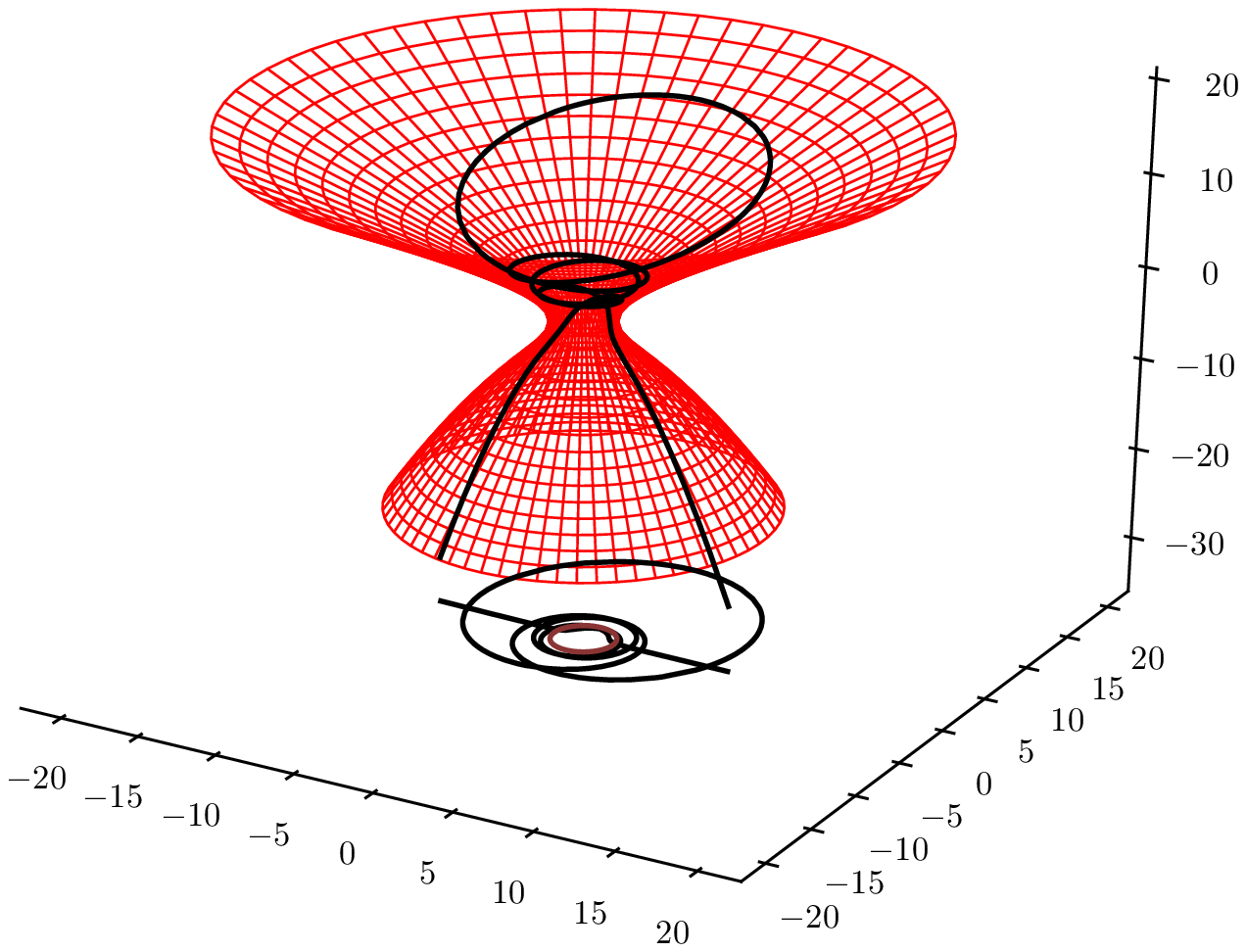}}
\subfloat[\RN{4}-orbits, $\bar{\epsilon}^2 = 2.00,\, \tilde{L} = 4.00,\, \bar{\alpha}_+ = 0.85,\, \zeta = 0$]{\label{Orbits:Fig:Orbits1:4}\includegraphics[width=0.45\linewidth]{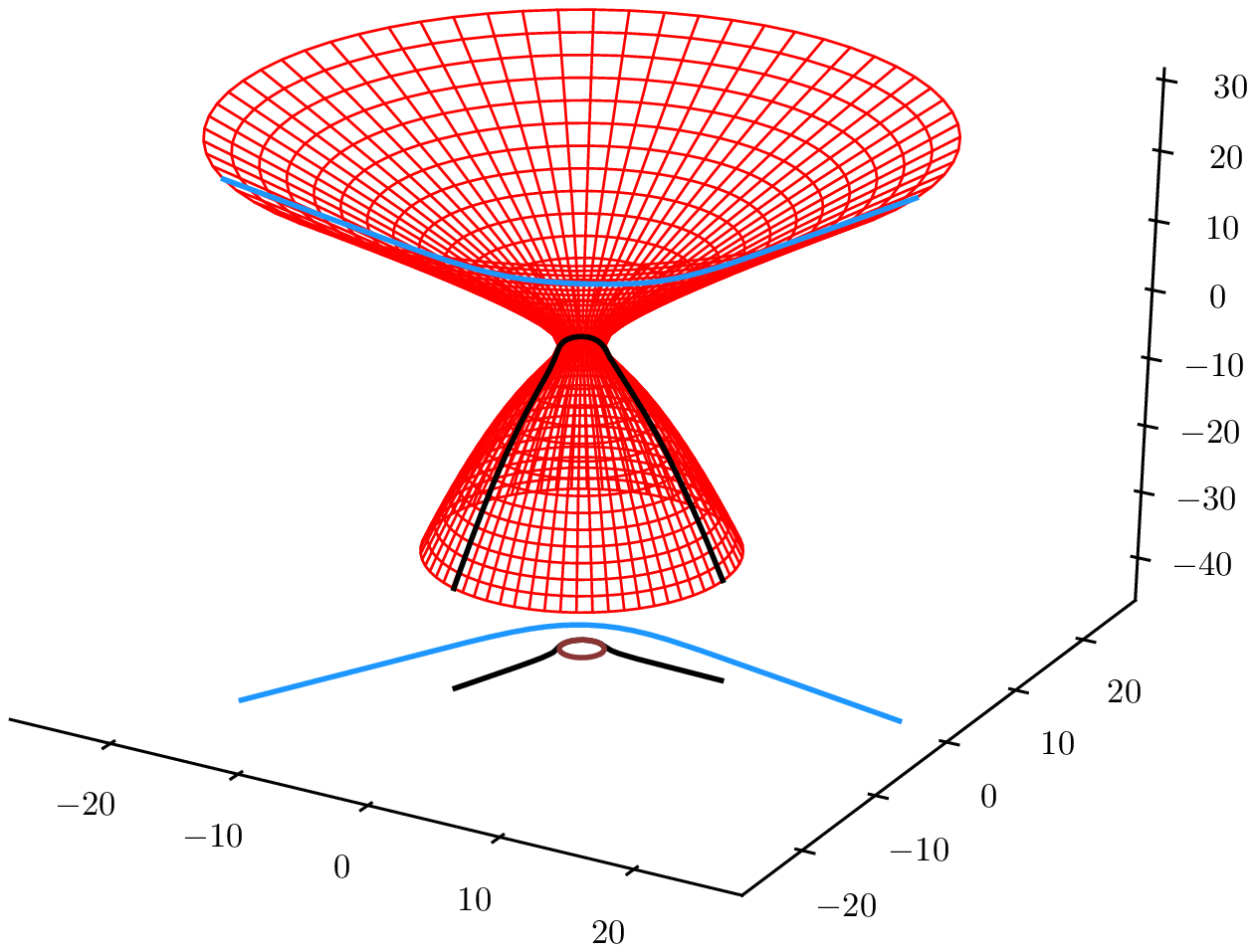}} \\
\subfloat[\RN{1}-orbit (TO), $\bar{\epsilon}^2 = 1.27,\, \tilde{L} = 3.00,\, \bar{\alpha}_+ = 0.30,\, \zeta = 1$]{\label{Orbits:Fig:Orbits1:5}\includegraphics[width=0.42\linewidth]{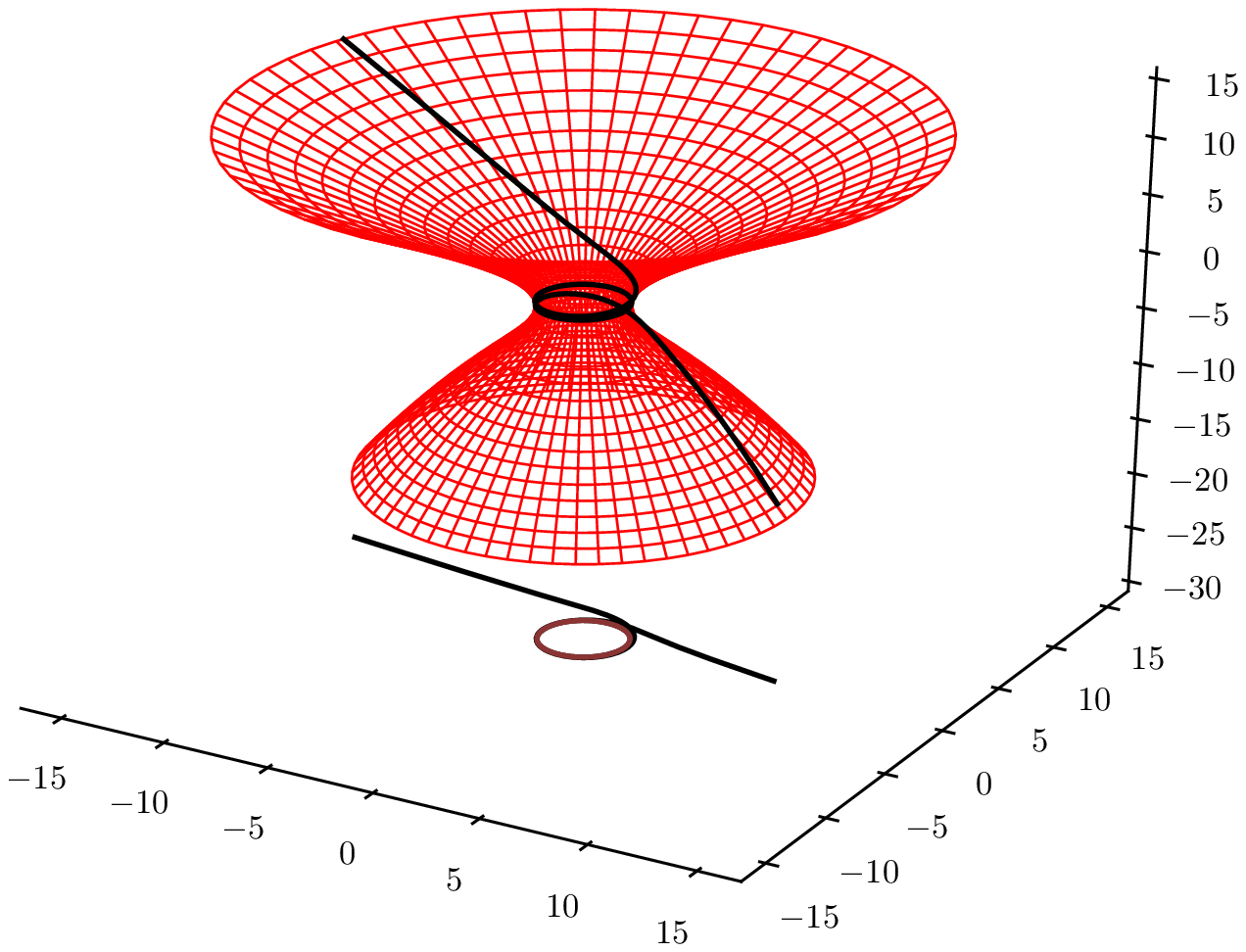}}
\subfloat[\RN{1}-orbit (TO), $\bar{\epsilon}^2 = 3.45,\, \tilde{L} = 3.00,\, \bar{\alpha}_+ = 0.85,\, \zeta = -1$]{\label{Orbits:Fig:Orbits1:6}\includegraphics[width=0.42\linewidth]{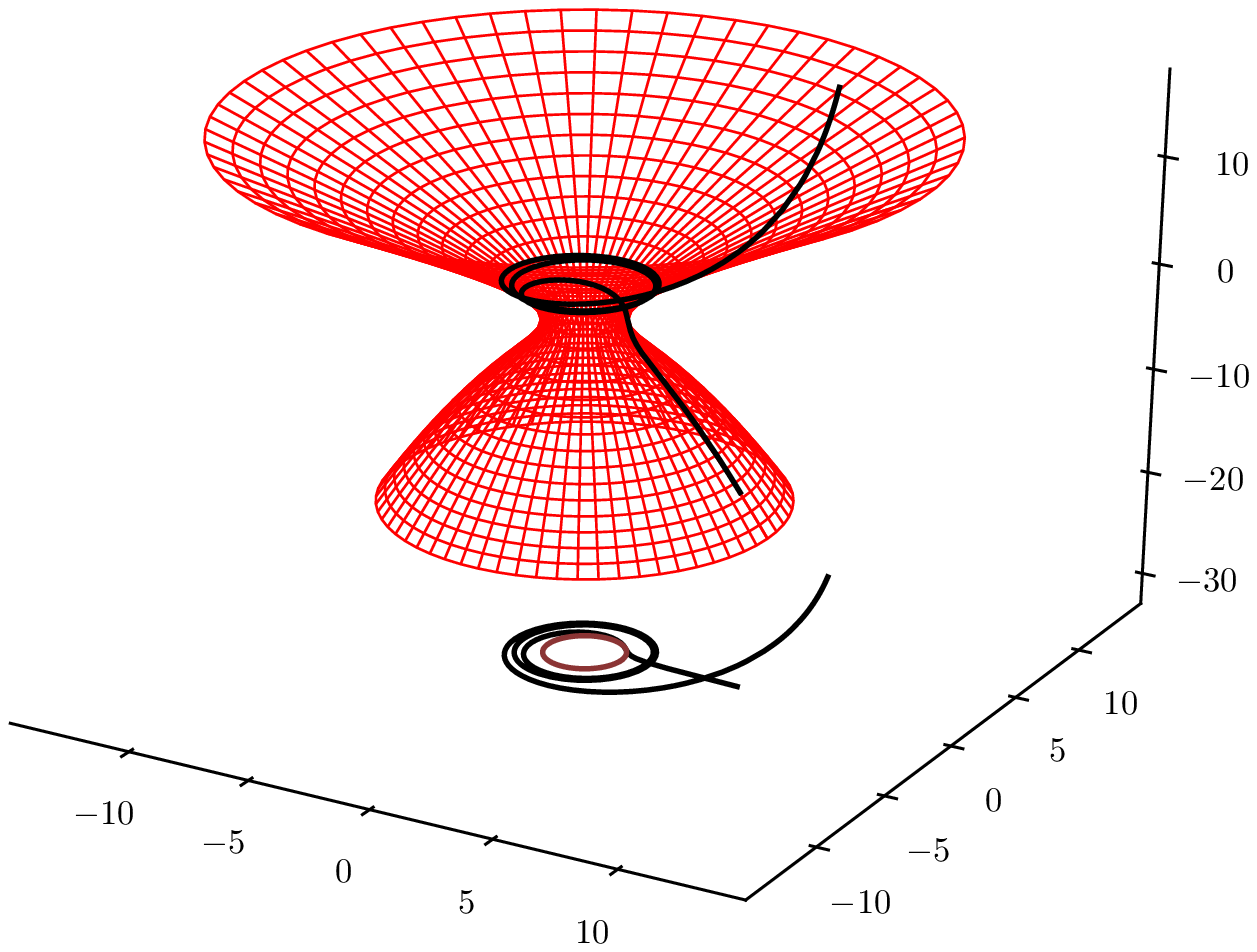}}
\caption{Orbits for various parameter sets calculated with the solution of $(d\bar{r}/d\phi)$. On the bottom of the figures the orbits are projected into the $X-Y$-plane, where the red inner circle represents the throat $\tilde{\rho}_T$. The light-blue line in \autoref{Orbits:Fig:Orbits1:4}represents the second geodesic possible for this orbit type.}
\label{Orbits:Fig:Orbits1}
\end{figure}

In the pictures of the isometrically embedding of the wormhole manifolds in \autoref{Orbits:Fig:WormholeManifold} the upper cone represents a part of the world with coordinate range $\bar{r} \in [\bar{r}_T,\, \infty]$, while the lower cone represents a part of the world with $\bar{r} \in [0,\, \bar{r}_T]$. A zoom of the Figs. \ref{Orbits:Fig:WormholeManifold}a-c highlighting the throat region is shown in the Figs. \ref{Orbits:Fig:WormholeManifold}d-f.

In \autoref{Orbits:Fig:Orbits1} we exhibit a set of trajectories as solutions of the geodesic equations on top of the corresponding wormhole embeddings. In particular, we exhibit orbits for different types of geodesics. As discussed above, bound orbits are located only in the upper cone, i.e., in the world with coordinate range $\bar{r} \in [\bar{r}_T,\, \infty]$, and two world escape orbits always pass from the lower cone to the upper cone, reach their turning point and return to the lower cone.

\section{Conclusion}
\label{Conclusion:Sec}
In this paper we have studied the geodesic motion in spacetimes describing traversable wormholes
supported by a massless conformally-coupled scalar field, found by Barcelo and Visser \cite{Barcelo:1999hq}. These static spherically symmetric wormholes connect two asymptotically flat worlds, which possess different physical properties. For instance, their masses, as read off from the asymptotic falloff of the metric, differ in both worlds \cite{Barcelo:1999hq}.

Nevertheless these wormhole spacetimes are interesting from a theoretical point of view, because they arise in General Relativity and do not need any exotic type of matter. Instead, the NEC violation results from the conformal coupling of the ordinary massless scalar field. Moreover, from a quantum field theory point of view the associated new improved energy-momentum tensor has finite matrix elements in the sense, that they are cutoff independent at large cutoff.

Here we presented the analytical solutions for geodesic motion in these wormhole spacetime. Restricting our discussion to $(d\bar{r}/d\phi)$ and $(d\bar{r}/dt)$, we obtained the solutions in terms of the Weierstra\ss \ $\wp$-, $\sigma$- and $\zeta$-functions. We also classified all possible orbits for timelike, lightlike and spacelike geodesics.

For timelike and spacelike motion the effective potential for the particle motion consists of a monotonic gravitational part and a non-monotonic centrifugal part, while for lightlike motion only the centrifugal part is present. Consequently, the classification depends on the amount of angular momentum $\tilde{L}$ of a particle. For timelike geodesics there are certain characteristic values $\tilde{L}_\text{crit}$ and $\tilde{L}_\text{swap}$ where the possible orbit types change. For spacelike geodesics there is in addition $\tilde{L}_\text{zero}$; however, only $\tilde{L}_\text{zero}$ matters when restricting to real energies. 

Stable bound orbits are only possible for timelike geodesics. They exist only in the upper world,where the gravitational potential is larger, and only when the angular momentum exceeds $\tilde{L}_\text{crit}$. Bound orbits can never cross the wormhole throat. (For spacelike geodesics such bound orbits would only be possible when considering particles with imaginary energy.)

Lightlike geodesics but also spacelike geodesics, whose angular momentum exceeds $\tilde{L}_\text{zero}$, have only transit orbits and two world escape orbits. However, whenever a maximum is present in the effective potential for timelike, lightlike, and spacelike orbits, unstable spherical orbits are possible as well. This means, in particular, that for any finite angular momentum there are unstable spherical light orbits. Thus these spacetimes possess a photosphere, implying the presence of a shadow, analogous to the case of other types of wormholes \cite{Bambi:2013nla,Nedkova:2013msa,Ohgami:2015nra,Shaikh:2018kfv}.

We note, that the full set of analytic solutions of the geodesic equations have also been obtained for the static, spherically symmetric Ellis (or Bronnikov-Ellis) wormhole \cite{Muller:2008zza}.However, this wormhole spacetime leads to a much simpler orbit classification. In particular, timelike, lightlike, and spacelike geodesics all lead to equal orbit types, and there are no stable bound orbits, making motion in this spacetime considerably less interesting \cite{Muller:2008zza}. This changes when the Ellis wormhole is set into rotation. Then stable bound orbits arise \cite{Kleihaus:2014dla,Chew:2016epf}. Moreover, when in addition ordinary bosonic matter is added a very interesting lightring structure appears \cite{Hoffmann:2017vkf,Hoffmann:2018}.

The wormhole spacetimes considered here are highly asymmetrical. However, as suggested by Barcelo and Visser \cite{Barcelo:1999hq}, one could obtain traversable wormhole solutions with no asymmetry, by adding thin shells of ordinary matter and joining smoothly inner and outer regions. For such wormhole  geometries similar techniques as the ones employed here could be used to study their orbits (see e.g.~\cite{Kagramanova:2013mwv}).

\section{Acknowledgements}
\label{Ackn:Sec}
We gratefully acknowledge support by the DFG Research Training Group 1620 {\sl Models of Gravity} and by the COST Action GWverse CA16104. Burkhard Kleihaus gratefully acknowledges support from Fundamental Research in Natural Sciences by the Ministry of Education and Science of Kazakhstan.


\end{document}